\documentclass[english,journal,comsoc]{IEEEtran}
\usepackage[T1]{fontenc}
\usepackage[latin9]{inputenc}
\usepackage{array}
\usepackage{float}
\usepackage{multirow}
\usepackage{amsmath}
\usepackage{graphicx}

\makeatletter

\providecommand{\tabularnewline}{\\}
\floatstyle{ruled}
\newfloat{algorithm}{tbp}{loa}
\providecommand{\algorithmname}{Algorithm}
\floatname{algorithm}{\protect\algorithmname}

\usepackage[T1]{fontenc}
\ifCLASSINFOpdf
\else
\fi
\usepackage{amsmath}
\usepackage[cmintegrals]{newtxmath}
\makeatother

\usepackage{babel}
\begin{document}

\title{Harmonic-summing Module of SKA on FPGA \textendash{} Optimising
the Irregular Memory Accesses}

\author{Haomiao Wang, Prabu Thiagaraj, and Oliver Sinnen}
\maketitle
\begin{abstract}
The Square Kilometre Array (SKA), which will be the world's largest
radio telescope, will enhance and boost a large number of science
projects, including the search for pulsars. The frequency domain acceleration
search is an efficient approach to search for binary pulsars. A significant
part of it is the harmonic-summing module, which is the research subject
of this paper. Most of the operations in the harmonic-summing module
are relatively cheap operations for FPGAs. The main challenge is the
large number of point accesses to off-chip memory which are not consecutive
but irregular. Although harmonic-summing alone might not be targeted
for FPGA acceleration, it is a part of the pulsar search pipeline
that contains many other compute-intensive modules, which are efficiently
executed on FPGA. Hence having the harmonic-summing also on the FPGA
will avoid off-board communication, which could destroy other acceleration
benefits. Two types of harmonic-summing approaches are investigated
in this paper: 1) storing intermediate data in off-chip memory and
2) processing the input signals directly without storing. For the
second type, two approaches of caching data are proposed and evaluated:
1) preloading points that are frequently touched 2) preloading all
necessary points that are used to generate a chunk of output points.
OpenCL is adopted to implement the proposed approaches. In an extensive
experimental evaluation, the same OpenCL kernel codes are evaluated
on FPGA boards and GPU cards. Regarding the proposed preloading methods,
preloading all necessary points method while reordering the input
signals is faster than all the other methods. While in raw performance
a single FPGA board cannot compete with a GPU, in terms of energy
dissipation, GPU costs up to 2.6x times more energy than that of FPGAs
in executing the same NDRange kernels.
\end{abstract}


\IEEEpeerreviewmaketitle

\section{Introduction}

\IEEEPARstart{T}{he} Square Kilometre Array (SKA) is built to extend
our understanding of the Universe and ourselves and it will be the
world's largest radio telescope array when finished~\cite{dewdney2009square}.
A number of key science goals are targeted by the SKA~\cite{carilli2004science}
project and one of them is strong-field tests of gravity using pulsars,
which are highly magnetized rotating neutron stars. Since most pulsar
signals are weaker than white noise and their details are unknown,
a number of techniques are employed to search for different types
of pulsars over a wide range of searching scales (e.g. sky coverage,
frequency, bandwidth, and integration time)~\cite{ransom2002fourier}.
The enormous signal rate of the SKA makes an efficient solution only
using general processors to complete the searching tasks in the given
time period extremely difficult.

Taking the high-performance computing ability, power consumption,
and flexibility into consideration, the field-programmable gate array
(FPGA) seems to be an ideal device to accelerate the Central Signal
Processor (CSP) of the SKA project. The SKA stage 1 (SKA1) project
plans to adopt high-end FPGAs to accelerate part of the function modules
in the CSP regarding pulsar search such as frequency domain acceleration
search. However, the general hardware description language (HDL, e.g.
Verilog HDL and VHDL) based development process makes it hard to achieve
fast prototyping design and design space exploration. Additionally,
developers of an internationally distributed team, including non-hardware
experts, would need to understand the hardware structure of FPGA devices.

To address these problems, we employed a high-level approach by using
a high-level language compared to HDL. In this paper, we take a pulsar
search module called harmonic-summing as a case study. The harmonic-summing
module is a part of the Fourier domain acceleration search (FDAS)
module that contains a compute-intensive module. The compute-intensive
module performs very well on FPGAs~\cite{putnam2014reconfigurable},
so in order to avoid unnecessary data transfer, it is important to
have the harmonic-summing module on the FPGA. The main feature of
the harmonic-summing module is that the access to the input signals
is irregular and this affects the hardware accelerator in achieving
high-performance computing. We investigate a number of methods and
architectures to optimise the irregular memory accesses of the harmonic-summing
module and using Open Computing Language (OpenCL) for the prototype
design. The main contributions are as follows:
\begin{enumerate}
\item $Reducing\,Intermediate\,Data\,Accesses$: The straight-forward and
proposed approaches for the harmonic-summing module are investigated
and designed. The proposed approach reduces the total number of off-chip
memory accesses by changing the processing order and storing the intermediate
data in on-chip memory.
\item $Preloading\,Data$: Based on the proposed approach, two preloading
data methods are investigated by: 1) loading points with high touch
frequency and 2) loading necessary points that are needed to calculate
a block of points. Both these methods preload data to on-chip memory
before processing and further reduce the total amount of off-chip
memory accesses. 
\item $Reordering\,Input$: Based on the preloading necessary points method,
we investigate reordering the input points to improve the memory access
speed. After reordering the input, the data needed for each work group
are from consecutive addresses and they can be streamed to the FPGA
from off-chip memory.
\item $Across\,Device\,Evaluation$: The proposed methods are implemented
on FPGA using OpenCL. We adjust and port the implementations to different
devices and evaluate on different series of FPGAs, general-purpose
graphics processing units (GPGPUs) and CPUs for comparison.
\end{enumerate}
The rest of the paper is organized as follows. Section~\ref{sec:Related-Work}
gives related work on optimising irregular memory accesses and high-level
tools for developing for FPGAs. Section~\ref{sec:Harmonic-summing-Module-and}
provides the details of the harmonic-summing module and the design
goals. In Section~\ref{sec:OpenCL-based-Architecture}, two approaches
of OpenCL-based designs of the harmonic-summing module are proposed
and compared. Section~\ref{sec:Evaluation} presents the evaluation
and results are discussed. Finally, the conclusions are given in Section~\ref{sec:Conclusions}. 

\section{\label{sec:Related-Work}Related Work}

\subsection{Irregular Memory Access Optimisation }

In hardware-based high-performance computing, the efficiency of data
transfer between the accelerator and the memory system is an important
factor. A large amount of research has been done to improve the memory
access efficiency for accelerators such as GPGPUs~\cite{jang2011exploiting}
and FPGAs.

For some applications, the accesses to memory are irregular that limits
the performance of the accelerator, and this problem has been well-studied~\cite{hiba2012memory}.
For most applications with irregular memory access, there are mainly
two types of optimisation techniques: 1) reducing the number of accesses
and 2) scheduling as many accesses in parallel\cite{weinhardt1999memory}.
These two methods can be applied to various platforms such as FPGAs~\cite{weinhardt2001memory}.
For some graph computation problems in~\cite{wang2015addressing},
an on-chip distributed off-chip shard memory architecture with high-performance
shuffle network was investigated and the intermediate buffers were
reduced to save off-chip memory bandwidth. In~\cite{yang2013optimizing},
prefetching is researched to reduce the number of memory accesses.
In \cite{jain2013linearizing}, an irregular stream buffer~(ISB)
that targets the irregular sequences of temporally correlated memory
references is proposed. Data and computation reordering is employed
in \cite{mellor2001improving} to improve memory hierarchy performance.
Besides these approaches, many compilers focus on irregular memory
access such as ROCCC~\cite{halstead2011exploring} for FPGAs and
Sparse matrix-vector multiplication(SMVM)~\cite{fowers2014high}. 

Regarding the optimisation of two-dimensional harmonic summing calculations
done in this research, we are not aware of any prior work which investigating
it on a large-scale, especially in the context of acceleration devices
such as GPUs and FPGAs.

\subsection{FPGA as an Accelerator}

High-end FPGAs have been widely adopted as accelerators in many commercial
applications and research areas such as high-frequency trading~\cite{leber2011high}
and cloud computing~\cite{eguro2012fpgas}. Because of the outstanding
energy-efficient performance over GPGPU devices, Microsoft applied
high-end FPGAs in their data centers~\cite{putnam2014reconfigurable},
and FPGA-based accelerators appear in other cloud data centers as
well~\cite{tarafdar2017enabling}. Several science projects of different
areas such as SKA~\cite{wang2016high}, CERN~\cite{sridharan2016accelerating},
and DNA sequence analysis~\cite{huang2017hardware} exist that employ
a large number of FPGA devices for acceleration, connected through
the PCI Express (PCIe) bus or Ethernet cable. 

Besides these, FPGAs are widely employed in radio astronomy projects
as accelerators. In~\cite{de2007radio}, hundreds of Xilinx Virtex-4
FPGAs are used to implement the correlator of the SKAMP project. In~\cite{sanchez2005digital},
FPGA platforms are employed to accelerate digital channelised receivers.
The Berkeley CASPER group, MeerKAT, and NRAO released an FPGA-based
acceleration device for implementing the FX correlator for radio telescope
array~\cite{parsons2009digital}.

\subsection{High-level Synthesis}

One barrier of employing FPGAs as accelerators are the usual use of
the HDL-based development process that makes the time-to-market longer
than GPGPUs and multi-core processors. To address this, many high-level
synthesis tools have been released. Two primary FPGA vendors, Intel
and Xilinx, provide developers with their high-level tools. Intel
released several high-level development tools such as high-level synthesis
(HLS) compiler, which supports C++ based development, and FPGA SDK
for FPGA, which supports OpenCL~\cite{chen2012invited,czajkowski2012opencl}
based development. Xilinx provides two main tools: 1) high-level synthesis
of C/C++ and SystemC and 2) SDAccel that supports OpenCL. Besides
these official tools, there are several open source high-level synthesis
tools such as LegUp~\cite{canis2011legup}. 

\subsubsection{OpenCL for Intel FPGA}

OpenCL is an open parallel programming language. The main advantage
of OpenCL is that it is compatible with different types of acceleration
devices such as GPGPUs, CPUs, and FPGAs. Intel released a dedicated
FPGA development tool using OpenCL, which is called Intel FPGA SDK
for OpenCL (AOCL). An FPGA-based OpenCL application is divided into
two parts: the host programs and the kernels for devices. The host
program is written in C/C++. Before launching an OpenCL kernel in
the host program, the arguments of it are set, and all necessary data
are sent to the off-chip memory of FPGA devices through PCIe bus.
OpenCL classifies memory into two types local memory and global memory,
with the understanding that access to local memory is faster than
global, but sharing is limited. For OpenCL on an FPGA local memory
corresponds to on-chip memory such as BRAM and global memory corresponds
to off-chip memory such as DDR3 on the FPGA board. In this research,
the Intel FPGAs are adopted to implement the harmonic-summing module,
so the optimisation syntax and techniques that are mentioned in this
paper are targeting Intel FPGAs and AOCL.

\subsubsection{Single Work-item and NDRange Kernels}

NDRange is an important attribute of an OpenCL kernel that represents
its index space. Based on OpenCL 1.0~\cite{khronosopencl}, it contains
three integer values, where each value specifies the extent of the
index space in a dimension. The FPGA-based OpenCL kernels can be classified
into two types based on their NDRange sizes: single work-item kernel
and NDRange kernel. For the single work-item kernel, its NDRange size
is (1,1,1), which means the index space for all three dimensions are
one, resulting in a single work-group with one work-item. The kernel
code of a single work-item kernel looks more like C/C++ code than
that of NDRange kernels. However, some OpenCL-based optimisation attributes
are included within the kernel code. Generally, there is at least
one loop in a single work-item kernel and the number of iterations
equals to the global work size of the NDRange kernel. The ideal case
of the single work-item kernel is to launch one iteration of the outermost
loop per clock cycle, which is called loop pipelining. Regarding NDRange
kernels, its NDRange size is larger than (1,1,1) and the overall work
size has to be divided into small groups. In each small work group,
a small group of data is processed. The size of an NDRange kernel
is normally related to the details of a task. For example, if a two-dimension
NDRange kernel is designed to process an image with 256 points ($16\times16$),
its global work size can be set as (16,16,1). In our research, both
two kernel types are studied and the combination of single work-item
and NDRange kernels are investigated. 

\section{\label{sec:Harmonic-summing-Module-and}Harmonic-summing Module }

\begin{figure*}[t]
\begin{centering}
\includegraphics[bb=15bp 15bp 1070bp 300bp,clip,scale=0.45]{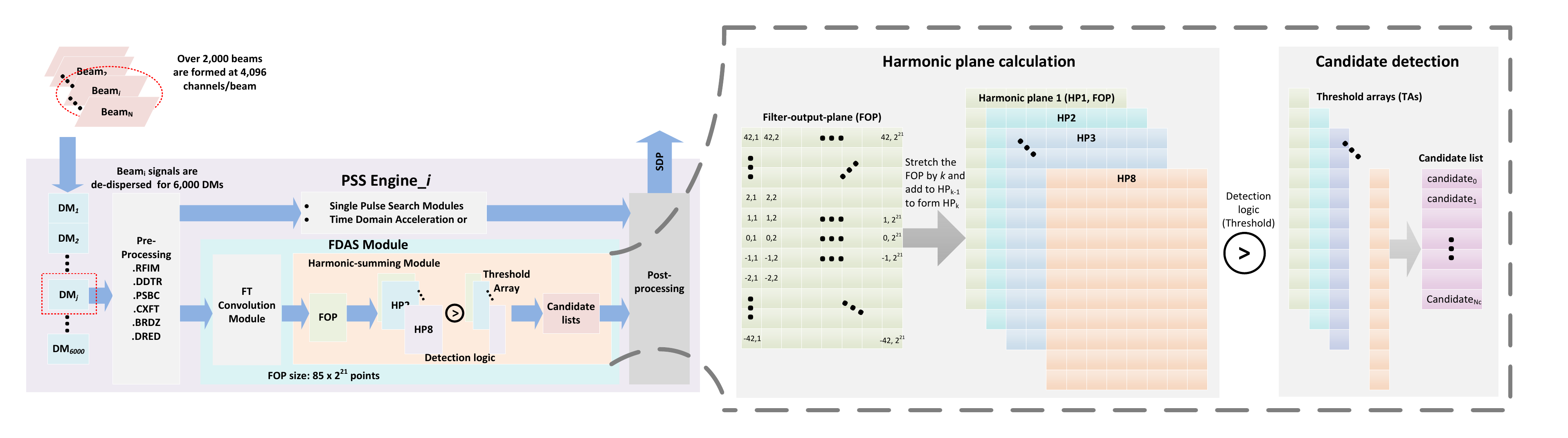}
\par\end{centering}
\caption{\label{fig:The-Processing-Flow}The processing flow of the Pulsar
Search Engine (PSS) of SKA1-MID CSP system and the details of harmonic-summing
module}
\end{figure*}

The harmonic-summing module is a part of the frequency domain acceleration
search~(FDAS) module~\cite{ransom2002fourier} of the pulsar search
engine~(PSS), whose details are depicted in Figure~\ref{fig:The-Processing-Flow}.
In the FT-based convolution module, the overlap-save algorithm~\cite{pavel2013algorithms}
is employed to process the input signals in the frequency domain and
the outputs are divided into chunks, several thousands values long.
The final output from the FT-based convolution module, which is also
the input of the harmonic-summing module, is called filter-output-plane
(FOP). The size of the FOP equals to $N_{temp}N_{chan}$, with $N_{temp}$
being the number of templates in the FT-based convolution and $N_{chan}$
being the number of channels $N_{chan}$. In essence, each template
is an FIR filter, and the FIR filter lengths of different templates
are different. The total $N_{temp}$ templates can be divided into
three groups, group one (index 1 to $(N_{temp}-1)/2$), group two
(index -1 to $-(N_{temp}-1)/2$), and the (unfiltered) input signals
(index 0, one-tap FIR filter). The number of channels is the same
as the length of input array of the FT convolution module. In our
previous work~\cite{wang2016fpga}, the FT convolution module has
been implemented in an FPGA using OpenCL. Based on current requirements,
an FOP contains $85\times2^{21}$ single precision floating-point~(SPF)
points, that is $N_{temp}=85$ and $N_{chan}=2^{21}$.

The harmonic-summing module (In Figure~\ref{fig:The-Processing-Flow}
(right)) consists of two parts: 1) harmonic plane calculation and
2) candidate detection. The task of the harmonic plane calculation
part is to generate $N_{hp}$ harmonic planes using the FOP. First,
the FOP is stretched by an integer $k$ to obtain the $k$th stretch
plane $SP_{k}$ , which is computed separately for template group
one and template group two by generating $N_{hp}$ stretch planes
with Equation~(\ref{eq:Sp}) 
\begin{equation}
SP_{k}(i,j)=SP_{1}(\left\lfloor \frac{i}{k}\right\rfloor ,\left\lfloor \frac{j}{k}\right\rfloor ),\,k=2,3,...N_{hp}\label{eq:Sp}
\end{equation}
 where $SP_{1}$ is the FOP and the ranges of $i$ and $j$ are $[-(N_{temp}-1)/2,\,(N_{temp}-1)/2]$
and $[0,\,N_{chan}-1]$, respectively. After all $N_{hp}-1$ stretch
planes are generated, the FOP and these $N_{hp}-1$ stretch planes
are progressively added to form $N_{hp}-1$ harmonic planes ($HP$s):

\begin{equation}
HP_{k}(i,\,j)=HP_{k-1}(i,\,j)+SP_{k}(i,\,j),\,k=2,3,...N_{hp}.\label{eq:sp_to_hp}
\end{equation}
It can be seen that the size of each $HP_{k}$ is the same as that
of the FOP. 

For the candidate detection, a threshold-detection logic is applied
and the potential candidates are recorded. For each harmonic plane,
a threshold array ($TA$) that contains $N_{temp}$ thresholds is
employed and one threshold corresponds to one row ($N_{chan}$ points)
of the harmonic plane. For example, $TA(k_{i})$ is the threshold
for the $i$th row of $HP_{k}$. In each harmonic plane, at most $N_{cand}$
candidates are stored and the maximum size of the candidate list for
each de-dispersion measure (DM) trail is $N_{hp}N_{cand}$. The output
from the candidate detection part is the candidate list and it will
be sent to the Fourier Domain Candidates optimisation (FDAO) module
for further processing (which is part of the post-processing in Figure~\ref{fig:The-Processing-Flow}).

Each candidate in the candidate list contains four elements:  periodicity,
orbits, pulse-width, and signal power of each detection. We use $\{F_{i},\,H_{i},\,B_{i,}\,A_{i}\}$
to represent the $i$th candidate in the candidate list, where $F_{i}$,
$H_{i}$, $B_{i}$, and $A_{i}$ are the index of filter, harmonic
plane, and bin and the amplitude of the $i$th element, respectively.
To minimize the data transfer bandwidth and save off-chip memory,
we use two 32-bit numbers, $CL_{i1}$ and $CL_{i2}$, to store the
$i$th candidate. For $F_{i}$, $H_{i}$, and $B_{i}$, the minimum
number of bits required is $\left\lceil log_{2}\left(*\right)\right\rceil $,
so the data sizes for them are 7-bit (85), 3-bit (8) and 21-bit ($2^{21}$),
respectively. These three factors can be combined together to form
one 32-bit integral $CL_{i1}$ by using the formula as follows: 
\[
CL_{i1}=F_{i}\times2^{24}+H_{i}\times2^{21}+B_{i}.
\]
In terms of the amplitude (spectral power), since the default data
type from the FT-based convolution module is SPF (Single Precision
Floating point, 32-bit), the same data type is maintained after the
harmonic-summing calculation, which means $CL_{i2}=A_{i}$.

The details of the harmonic summing algorithm are given in Algorithm~\ref{alg:General-Harmonic-summing-Algorit},
where the order of the three\texttt{ for }loops can be interchanged.
The basic parameters of the harmonic-summing module are shown in Table~\ref{tab:Harmonic-summing-Module-Paramete}.

\begin{table}
\caption{\label{tab:Harmonic-summing-Module-Paramete}Specification of the
harmonic-summing module}
\centering{}{\small{}}%
\begin{tabular}{|c|c|c|}
\hline 
{\small{}Parameter} & {\small{}Description} & {\small{}Value}\tabularnewline
\hline 
\hline 
{\small{}$N_{temp}$} & {\small{}Number of templates of the }FOP{\small{} (row)} & {\small{}$85$}\tabularnewline
\hline 
{\small{}$N_{chan}$} & {\small{}Number of channels of the }FOP{\small{} (column)} & {\small{}$2^{21}$}\tabularnewline
\hline 
{\small{}$N_{hp}$} & {\small{}Total number of harmonic planes} & {\small{}$8$}\tabularnewline
\hline 
{\small{}$N_{cand}$} & {\small{}Number of candidates per harmonic plane} & {\small{}$200$}\tabularnewline
\hline 
{\small{}$t_{limit}$} & {\small{}Computation time limit of each $DM$ trail} & {\small{}$88ms$}\tabularnewline
\hline 
\end{tabular}{\small \par}
\end{table}

\begin{algorithm}
\caption{\label{alg:General-Harmonic-summing-Algorit}General Harmonic-summing
Algorithm\textsc{ (SingleHP)}}

{\small{}$SP_{1}\leftarrow$(filter-output-plane)}{\small \par}

{\small{}$CL\leftarrow0$ \{initialize the detection output\}}{\small \par}

\textbf{\small{}for $k=1$ }{\small{}to $N_{hp}$ }\textbf{\small{}do}{\small \par}

{\small{}~~~~}\textbf{\small{}for $i=-(N_{temp}-1)/2$ }{\small{}to
$(N_{temp}-1)/2$ }\textbf{\small{}do}{\small \par}

{\small{}~~~~}\textbf{\small{}~~~~for $j=0$ }{\small{}to
$N_{chan}-1$ }\textbf{\small{}do}{\small \par}

{\small{}~~~~~~~~~~~~$SP_{k}(i,\,j)\leftarrow$stretch$(SP_{1},\,k,\,i,\,j)$
\{generate the value in stretched plane\}}{\small \par}

{\small{}~~~~~~~~~~~~$HP_{k}(i,\,j)\leftarrow HP_{k-1}(i,\,j)+SP_{k}(i,\,j)$
\{based on the stretched plane, generate the value in harmonic plane\}}{\small \par}

{\small{}~~~~~~~~~~~~$CL\leftarrow$append detection$[HP_{k}(i,\,j),\,TA(k,\,i)]$
\{threshold-detection logic to identify valid peak signals\}}{\small \par}

{\small{}~~~~~~~~}\textbf{\small{}end for}{\small \par}

{\small{}~~~~}\textbf{\small{}end for}{\small \par}

\textbf{\small{}end for}{\small \par}

{\small{}Candidate List $\leftarrow CL$}{\small \par}
\end{algorithm}

\section{\label{sec:Motivation}Proposed Methods}

The main problem for the harmonic summing module is the irregular
memory accesses of the harmonic plane calculation part and it limits
the data transfer efficiency. We consider two types of memory access
optimisation methods while designing the harmonic plane calculation
part: 1) increasing the off-chip memory bandwidth and 2) reducing
the number of off-chip memory accesses. Based on the number of processed
harmonic planes at a time, two approaches are investigated: the \textsc{SingleHP}
method (processing a single harmonic plane at a time) and the \textsc{MultipleHP}
method (processing multiple harmonic planes at a time). 

\subsection{\label{subsec:Design-Goals}Design Goals}

In designing the harmonic-summing module, we mainly consider the latency
and energy dissipation of calculating the harmonic planes and detecting
the candidates using high-end FPGAs. There are two major factors that
affect the execution latency and energy dissipation: 1) parallelisation
capacity of an FPGA and 2) data transfer rate between the FPGA and
off-chip memory. Most operations in the harmonic-summing module are
floating-point operations, however, they are inexpensive functions
such as floating-point additions and comparisons with a constant.
For high-end FPGAs, there are hundreds of DSP blocks (to implement
Floating point operations) and hundred thousand of logic elements
that can handle these operations effectively. 

\begin{figure}
\begin{centering}
\includegraphics[bb=24bp 20bp 670bp 460bp,clip,scale=0.35]{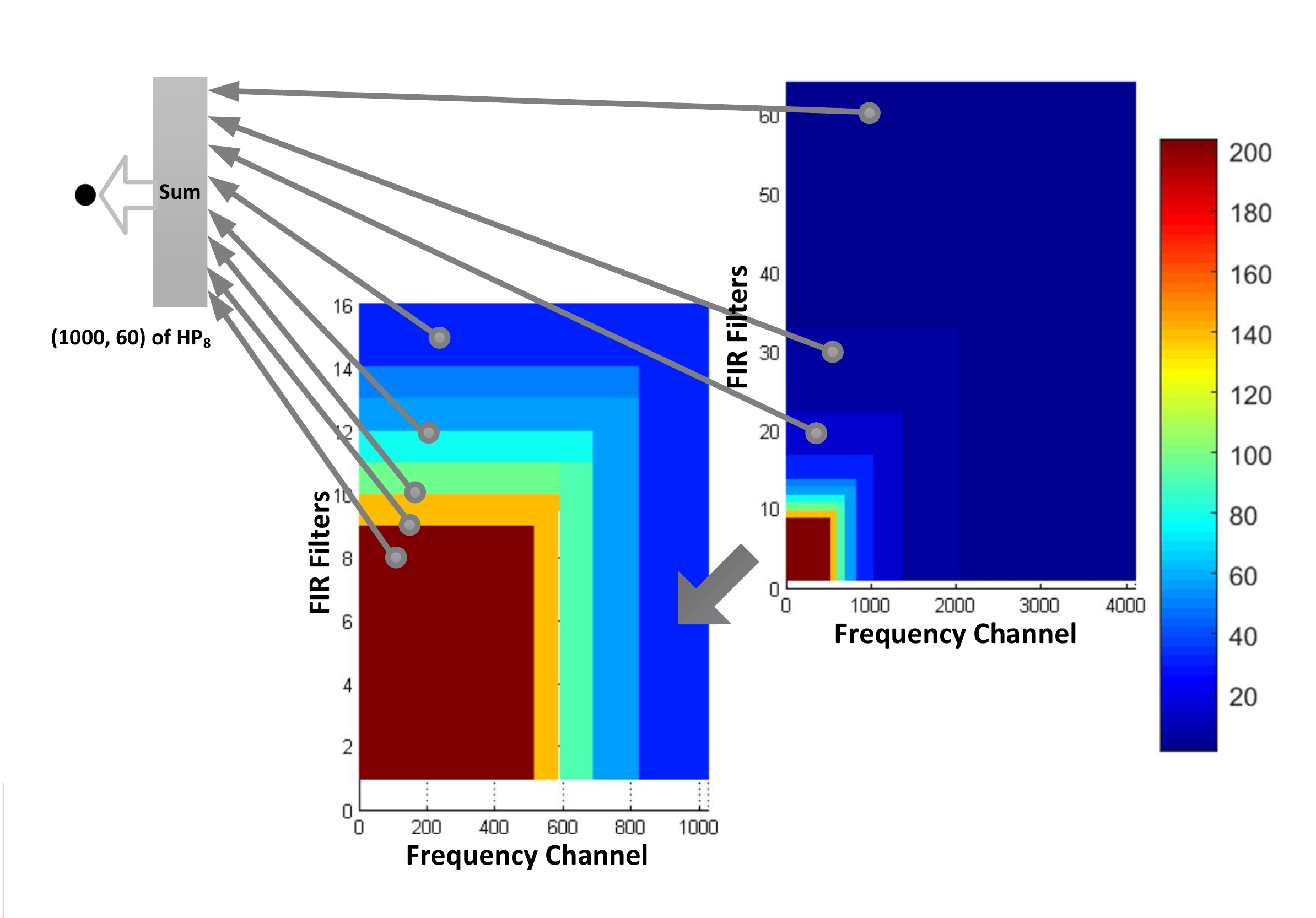}
\par\end{centering}
\caption{\label{fig:Touching-frequency-of}Touching frequency of each point
in the FOP and an example of calculating point $(1000,\,60)$ of $HP_{8}$ }
\end{figure}

In the harmonic plane calculation, the accesses to off-chip memory
is not consecutive but irregular due to the index calculations in
Equation~(\ref{eq:Sp}). Ideally, the data transfer bandwidth of
any design equals to the device's theoretical maximum bandwidth, however,
this cannot be achieved easily in the harmonic-summing module. Taking
a small size FOP ($64\times2^{12}$) as an example, the touching frequencies
of the FOP elements in calculating 7 harmonic planes are depicted
in Figure~\ref{fig:Touching-frequency-of}. For example, 8 points
from different positions are needed to calculate point $(1000,\,60)$
of $HP_{8}$. In this figure, the size of the deep red area is only
1.7\% of the whole FOP, however, each value is touched $204$ times.
The size of the high touching frequency area (zoomed-in area) is \textbf{$16\times2^{10}$}
and the sum of the touching times of this area is $73.4\%$ of the
overall touching times. It can be seen that the distribution of touching
frequency and memory access while calculating do exhibit a very complex
pattern. In this paper, we investigate a general design of the harmonic-summing
module with low latency, by optimising memory accesses.

The input to the harmonic-summing module, which is the FOP, is up
to $710MBytes$ under current requirements and it exceeds the on-chip
memory size of high-end FPGAs and other types of processors. Though
the FOP can be transferred to FPGAs through PCIe bus or Ethernet cable
in practice, it is assumed in this research that the FOP is stored
in off-chip memory before processing the harmonic-summing module (for
example as the output of the FT-convolution also executed on the FPGA
device~\cite{wang2016fpga}. 

In terms of the candidate detection of the harmonic summing module,
when there are more than $N_{cand}$ candidates detected in one harmonic
plane, the strategy of sorting candidates has not yet been settled
in the PSS sub-project. There are a number of plausible strategies
to select candidates, such as storing the largest $N_{cand}$ candidates
or first/last $N_{cand}$ candidates. 

Due to the lack of a settle requirement, and with the assumption that
there are usually less than $N_{cand}$ candidates (which can be tuned
by increasing thresholds), we investigate the methods of storing the
last $N_{cand}$ candidates. The FPGA device needs to go through all
the candidates from each harmonic plane. When there are less than
or equal to $N_{cand}$ candidates in one harmonic plane, all the
candidates will be recorded. Note that based on the method and process
order of harmonic plane calculation, the recorded last $N_{cand}$
candidates might vary between different approaches.

\subsection{\textsc{SingleHP}}

For the algorithm in Algorithm~\ref{alg:General-Harmonic-summing-Algorit},
the processor needs to calculate all harmonic planes individually.
 The \textsc{SingleHP} method is a straightforward implementation
of the harmonic-summing module. 

To calculate the points of the $k$th harmonic plane $HP_{k}$ ($k\geq2$),
points of the FOP and the $k-1$th harmonic plane $HP_{k-1}$ are
required. During processing, each generated point of $HP_{k}$ is
compared with a threshold. Since the FOP size, $N_{temp}N_{chan}$,
exceeds the on-chip memory of FPGA devices, the FOP and other generated
harmonic planes have to be stored in the off-chip memory of FPGA device. 

The accesses of loading points from $HP_{k-1}$ and storing points
to $HP_{k}$ are both in-order and of consecutive addresses. However,
the accesses of loading points from the FOP cannot be calculated as
a simple offset. So the data cannot be steamed between off-chip memory
and device while processing. To optimise the memory accesses of the
\textsc{SingleHP} method, the overall pipeline can be parallelised
to increase the off-chip memory bandwidth, and we use that in our
implementation.

\subsection{\textsc{\label{subsec:MultipleHP}MultipleHP}}

In the harmonic summing module, only the candidates are recorded for
further processing, it is unnecessary to store the data of all harmonic
planes in off-chip memory. To reduce the number of off-chip memory
accesses, we investigate the method to get rid of storing harmonic
planes except for the FOP.  If the points of the same index in multiple
or all $N_{hp}$ harmonic planes can be generated in parallel, these
points can be discarded directly after candidate detection. Without
storing the generated points back to off-chip memory, the number of
overall off-chip memory accesses can be halved.

By reordering the three\texttt{ for }loops in Algorithm~\ref{alg:General-Harmonic-summing-Algorit},
we obtain Algorithm~\ref{alg:mhp}, where the innermost\texttt{ for
}loop can be parallelised and the points are discarded after detection.

\begin{algorithm}
\caption{\label{alg:mhp}Multiple Harmonic-summing planes based method \textsc{(MultipleHP})}

{\small{}$SP_{1}\leftarrow$(filter-output-plane)}{\small \par}

{\small{}$CL\leftarrow0$ \{initialize the detection output\}}{\small \par}

\textbf{\small{}for}{\small{} }\textbf{\small{}$j=0$ }{\small{}to
$N_{chan}-1$}\textbf{\small{}do}{\small \par}

{\small{}~~~~}\textbf{\small{}for $i=-(N_{temp}-1)/2$ }{\small{}to
$(N_{temp}-1)/2$ }\textbf{\small{}do}{\small \par}

{\small{}~~~~}\textbf{\small{}~~~~for $k=1$ }{\small{}to
$N_{hp}$ }\textbf{\small{}do}{\small \par}

{\small{}~~~~~~~~~~~~$SP_{k}(i,\,j)\leftarrow$stretch$(SP_{1},\,k,\,i,\,j)$
\{generate the value in stretched plane\}}{\small \par}

{\small{}~~~~~~~~~~~~$HP_{k}(i,\,j)\leftarrow HP_{k-1}(i,\,j)+SP_{k}(i,\,j)$
\{based on the stretched plane, generate the value in harmonic plane\}}{\small \par}

{\small{}~~~~~~~~~~~~$CL\leftarrow$detection$[HP_{k}(i,\,j),\,TA(k,\,i)]$
\{threshold-detection logic to identify valid peak signals\}}{\small \par}

{\small{}~~~~~~~~}\textbf{\small{}end for}{\small \par}

{\small{}~~~~~~~~discard{[}$HP_{1}(i,\,j),\,$$HP_{2}(i,\,j),\,...,HP_{N_{hp}}(i,\,j)${]}\{discard
the point of same index after detection\}}{\small \par}

{\small{}~~~~}\textbf{\small{}end for}{\small \par}

\textbf{\small{}end for}{\small \par}

{\small{}Candidate List $\leftarrow CL$}{\small \par}
\end{algorithm}

To optimise the \textsc{MultipleHP} method by reducing the off-chip
memory accesses, part of the FOP can be loaded before calculating
a chunk of points of all harmonic planes. Two alternatives are proposed
and based on the loaded data, they can be distinguished as 1) high
touching frequency (by loading as many points as possible in the high
touching frequency area of the FOP) and 2) necessary points (by loading
points that are needed to calculate a chunk of points in all harmonic
planes such as one or several columns of all harmonic planes). For
the second method, an FOP reordering method is proposed below to increase
data transfer efficiency. Each of these three \textsc{MultipleHP}-based
methods adopted at least one type of memory accesses optimisation
method and the details of them are as follows.

\subsubsection{Preloading Points with High Touching Frequency}

To create and threshold test 8 consecutive harmonic planes, each point
with the highest touching frequency needs to be loaded over 200 times.
 If most points with high touching frequency can be preloaded, a
large number of load operations can be saved. To further reduce the
amount of off-chip memory accesses, part or all of the high touching
frequency points can be preloaded in on-chip memory. We use \textsc{MultipleHP-H}
to represent the preloading points with high touching frequency method.

\begin{figure}
\begin{centering}
\includegraphics[bb=0bp 10bp 504bp 250bp,clip,scale=0.5]{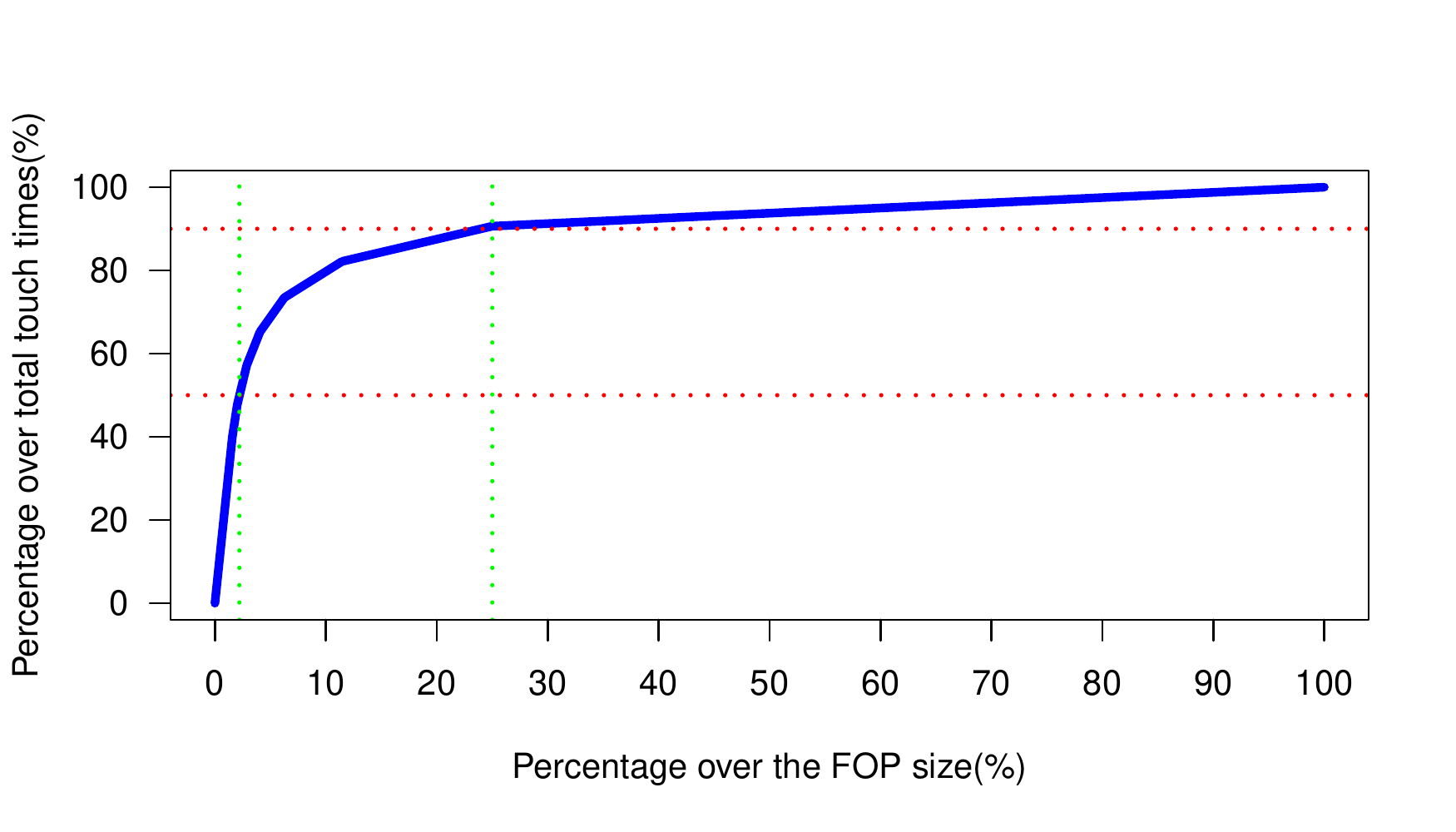}
\par\end{centering}
\caption{\label{fig:Relationship-between-the}Relationship between the size
of preloaded points and the reduced number of global memory accesses}
\end{figure}

The main factor of the \textsc{MultipleHP-H} method is the number
of preloaded high touching points $N_{MultipleHP-H-preld}$. If the
points in the FOP are sorted by touching times, the relationship between
the percentage of the FOP size and the percentage of overall touching
times is depicted in Figure~\ref{fig:Relationship-between-the}.
It can be seen that 2.2\% points in the FOP have about 50\% of overall
touching times and 25\% points have 90\% percent of overall touching
times.

\subsubsection{\label{subsec:Loading-Necessary-Points}Loading Necessary Points}

For the Na\text{\"i}ve \textsc{MultipleHP }method, calculating one
point with the same index of $N_{hp}$ harmonic planes, at most $N_{hp}$
points need to be loaded from the FOP. However, calculating a chunk
of points in all $N_{hp}$ harmonic planes need less than $N_{hp}$
times the number of points. Take one column with $N_{temp}$ points
as an example, it needs $N_{temp}$ points for $HP_{1}$, however,
$2\left\lceil (N_{temp}-1)/2\right\rceil +1$ points for $HP_{2}$,
$2\left\lceil (N_{temp}-1)/3\right\rceil +1$ points for $HP_{3}$
and so on. To save loading operations, the harmonic plane calculation
task can be decomposed into a number of work-groups. The task of each
work group is to generate a number of columns $N_{MultipleHP-N-col}$
of all $N_{hp}$ harmonic planes, where each column has $N_{temp}$
points. In a pipeline, the loading part of a work-group can overlap
with the computing part of the previous work-group. We use \textsc{MultipleHP-N}
to represent the loading necessary points method.

For the\textsc{ MultipleHP-N} method, $N_{MultipleHP-N-col}$ is an
important factor that affects the reduced off-chip memory accesses.
Assuming the task for each work-group is to generate one column ($N_{MultipleHP-N-col}=1$)
of $N_{hp}$ harmonic planes ($N_{hp}N_{temp}$ points in total) and
the maximum needed data is 
\[
2\sum_{i=1}^{N_{hp}}\left\lceil \frac{N_{temp}-1}{2i}\right\rceil +N_{hp}
\]
instead of $N_{hp}N_{temp}$ points. When $N_{MultipleHP-N-col}$
is larger than one, more off-chip memory accesses can be reduced.
However, the amount of data needed for the same harmonic plane varies
based on the column index. For example, if the work-group generates
eight columns ($N_{MultipleHP-N-col}=8$) of all harmonic planes,
the data needed to generate the $3$rd harmonic plane are 3 to 4 columns
of the FOP. To guarantee that the amount of data loaded for each work-group
is a constant (which is needed for efficient pipelining), the maximum
number of points for each harmonic plane is chosen.

\begin{figure}
\begin{centering}
\includegraphics[bb=0bp 15bp 504bp 240bp,clip,scale=0.5]{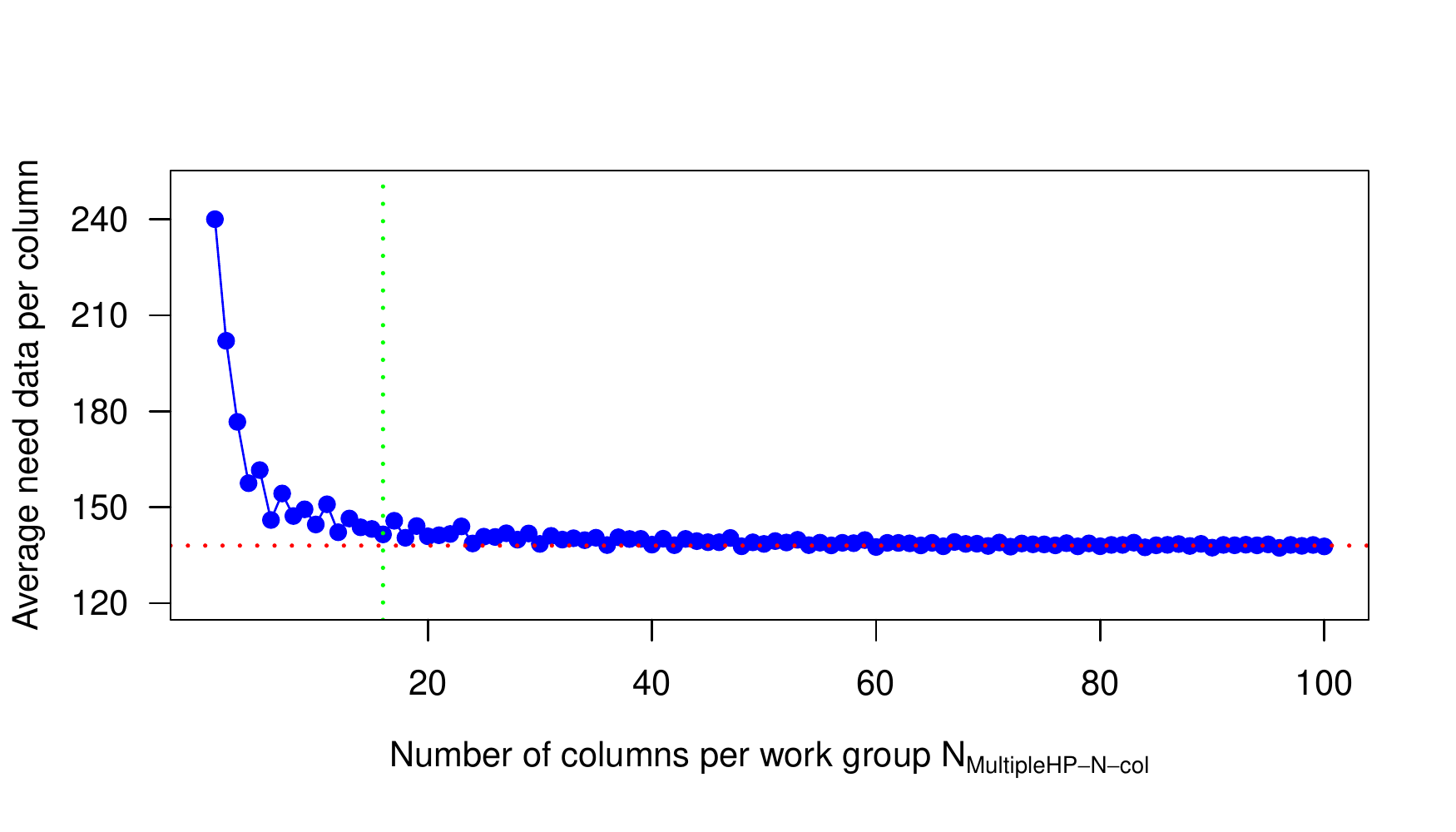}
\par\end{centering}
\caption{\label{fig:Relationship-between-columns/wor-1}Relationship between
columns per work group and the number of points per column for the
\textsc{MultipleHP-N} method }
\end{figure}

In this case, when the $N_{MultipleHP-N-col}$ is specified, the needed
number of columns for each harmonic plane can be listed and then the
number of needed points for $N_{MultipleHP-N-col}$ columns can be
calculated. Based on the number of overall needed points, the average
needed points per column for a work-group is plotted in Figure~\ref{fig:Relationship-between-columns/wor-1}.
It can be seen that the average amount drops fast when the value of
$N_{MultipleHP-N-col}$ is smaller than 16 (green dot line) and it
decreases slightly toward 64 (red dot line) as the $N_{MultipleHP-N-col}$
increases. Besides these, the larger the $N_{MultipleHP-N-col}$,
the larger space it needs in the on-chip memory. If the $N_{MultipleHP-N-col}$
is too large, the on-chip memory size might limit the $N_{MultipleHP-N-col}$.
As a consequence, it is unnecessary to assign tens or hundreds of
columns to a work group.

\subsubsection{\label{subsec:Reordering-the-FOP}Reordering the FOP}

Comparing with the Na\text{\"i}ve \textsc{MultipleHP }method, the
\textsc{MultipleHP-N} method can further reduce the total amount of
off-chip memory accesses. However, the points needed for each work-group
are from at least $N_{hp}$ blocks in FOP and they are from non-consecutive
addresses. Thus, the points for each work-group cannot be streamed
between off-chip memory and FPGA device. 

To optimise the off-chip memory bandwidth of the \textsc{MultipleHP}\textbf{\textsc{-}}\textsc{N}
method, we propose the \textsc{MultipleHP}\textbf{\textsc{-}}\textsc{R}
method which reorders the FOP to form the reordered FOP~(\textsc{r}FOP).
After reordering, the needed points to calculate $N_{MultipleHP-R-col}$
columns of all harmonic planes are from consecutive addresses that
can be streamed to the FPGA while processing. However, the size of
the reordered FOP is larger than the standard FOP size. Theoretically,
the number of rows in the reordered \textsc{r}FOP is increased from
$N_{temp}$ to the average needed points per column in Figure~\ref{fig:Relationship-between-columns/wor-1}.
Take the $N_{MultipleHP-R-col}=16$ as an example, the smallest average
needed points per column is 141.5 that makes the size of \textsc{r}FOP
at least 1.66x times larger than the original FOP size. It can be
seen that the larger the $N_{MultipleHP-R-col}$, the smaller the
relative size of \textsc{r}FOP. The details of \textsc{r}FOP generation
and optimisation are discussed in Section~\ref{subsec:-MultipleHP-R}
The latency of extra data transfer and FOP reordering have to be considered
in the evaluation of the \textsc{MultipleHP}\textbf{\textsc{-}}\textsc{R}
method.

\section{\label{sec:OpenCL-based-Architecture} Architecture and Optimisation}

In this section, we investigate the architecture of the proposed methods
and employ OpenCL as the high-level language, whose kernels can be
executed on both FPGAs and GPUs. Having that said, since the goal
is to evaluate FPGA performance, the optimisation techniques and syntax
are dedicated to FPGAs.

\subsection{\textsc{SingleHP} kernel}

\begin{figure}
\begin{centering}
\includegraphics[bb=15bp 15bp 455bp 235bp,clip,scale=0.55]{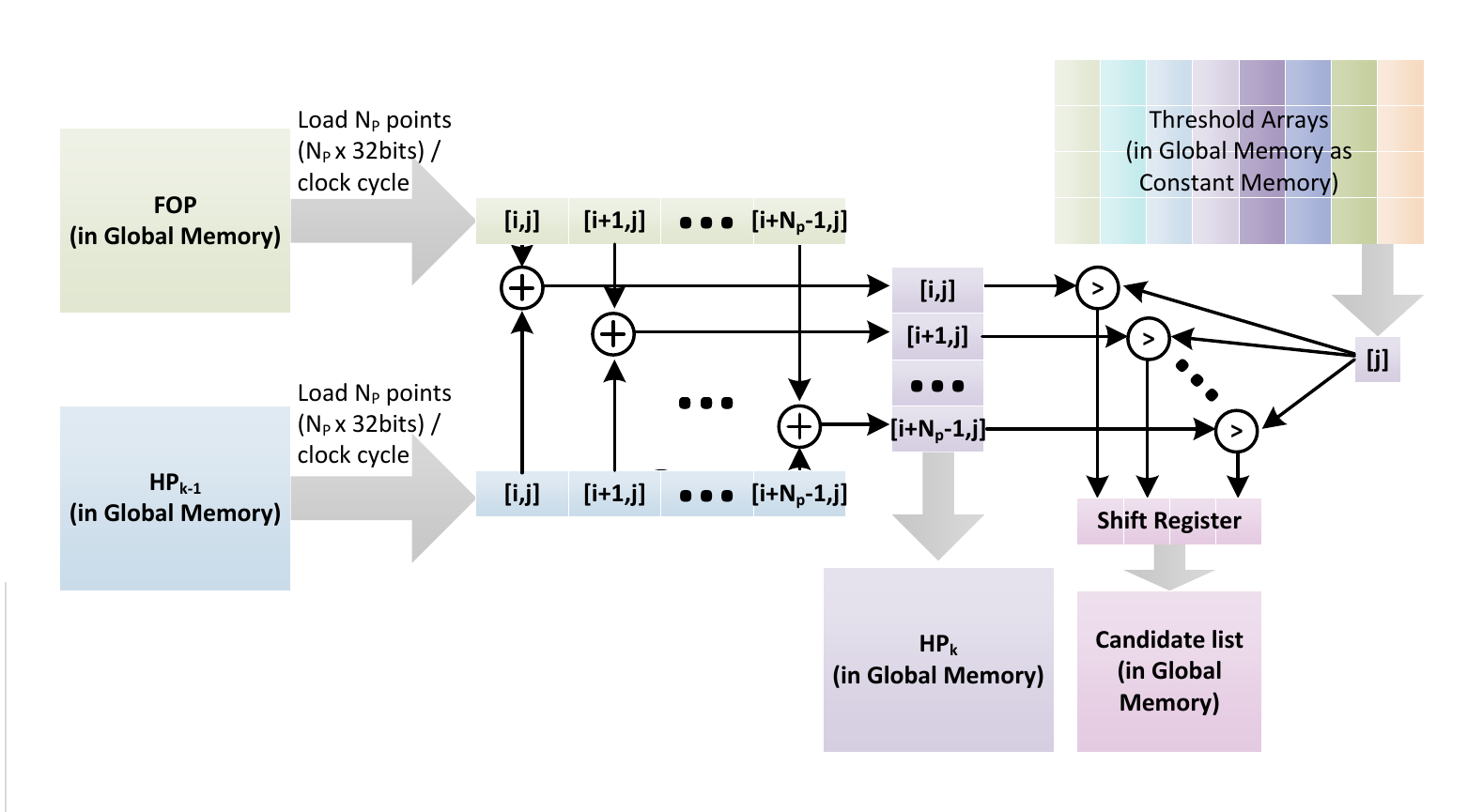}
\par\end{centering}
\caption{\label{fig:Plane-by-Plane}Architecture of the \textsc{SingleHP} kernel}
\end{figure}

The basic structure of the \textsc{SingleHP} kernel while processing
the $k$th harmonic plane $HP_{k}$ is depicted in Figure~\ref{fig:Plane-by-Plane},
where $N_{paral}$ is the parallelisation factor that is restricted
by global memory (off-chip memory in this research) bandwidth (GMB)
and the logic resources of the FPGA. One optimisation goal for the
\textsc{SingleHP} kernel is to find the maximum parallelisation factor
$N_{paral-max}$ that leads to a required GMB which equals the physical
off-chip memory bandwidth of a specific device. 

The FOP, $HP_{k-1}$, $HP_{k}$, candidate list and $TA$ are all
stored in global memory before launching the kernel. When the kernel
is launched, $N_{paral}$ points from $HP_{k-1}$ and $\left\lceil \frac{N_{hp}}{k}\right\rceil $
points from FOP are loaded per clock cycle. These points are summed,
according to Equation~(\ref{eq:Sp}) and Equation~(\ref{eq:sp_to_hp}),
to calculate $N_{paral}$ points of $HP_{k}$. The generated $N_{paral}$
points are compared with the corresponding thresholds and detected
candidates are saved in a shift register or local memory (on-chip
memory in this research) of length $N_{cand}$, until all FOP points
have been processed. Then these $N_{paral}$ points overwrite the
values at the same address of $HP_{k-1}$. 

In OpenCL, both single work-item and NDRange kernel types can be adopted
to implement the \textsc{SingleHP} kernel. 

To parallelise $N_{paral}$ points in a single work-item kernel, we
partially unroll the outermost loop by a factor of $N_{paral}$ (\texttt{\small{}\#pragma
unroll $N_{paral}$}). Before partially unrolling the outermost loop,
the innermost loops are completely unrolled to achieve loop pipelining. 

For the NDRange kernel, kernel vectorisation (\texttt{\small{}num\_simd\_work\_items($N_{paral}$)})
and compute unit replication (\texttt{\small{}num\_compute\_units($N_{paral}$)})
techniques can be employed to parallelise the kernel. Note that detected
$N_{cand}$ candidates might be different for the vectorized and replicated
kernels, under the condition that the threshold has been triggered
more than $N_{cand}$ times. As only the last $N_{cand}$ candidates
are stored, different parallelisation result in different execution
orders and hence candidates.

The \textsc{SingleHP} kernel can be implemented as a generic kernel
that needs to be launched $N_{hp}$ times (multiple launches) or a
specific kernel that only needs to be launched once (single launch)
to generate the candidate list of $H_{hp}$ harmonic planes. The overhead
of launching a kernel such as setting kernel arguments will affect
the overall latency, especially when the kernel execution latency
is short. So the kernel launch time is an important factor for the
\textsc{SingleHP} kernel. Multiple launches provide more flexibility
than the single launch \textsc{SingleHP} kernel, as it can be used
for any harmonic plane configuration. Both single and multiple launches
kernels are evaluated in Section \ref{sec:Evaluation}.

\subsection{\label{subsec:One-FOP}\textsc{MultipleHP }Methods based Kernels}

Although parallelising the \textsc{SingleHP} kernel can shorten kernel
execution latency by increasing GMB, the total amount of global memory
accesses (GMA) is not reduced. The main advantage of the\textsc{ MultipleHP}
method is the reduction of the required GMA by processing multiple
harmonic planes at the same time. A number of optimisation techniques
are investigated for the \textsc{MultipleHP}-based methods in the
following.

\subsubsection{Na\text{\"i}ve \textsc{MultipleHP}}

\begin{figure}
\begin{centering}
\includegraphics[bb=20bp 15bp 515bp 290bp,clip,scale=0.5]{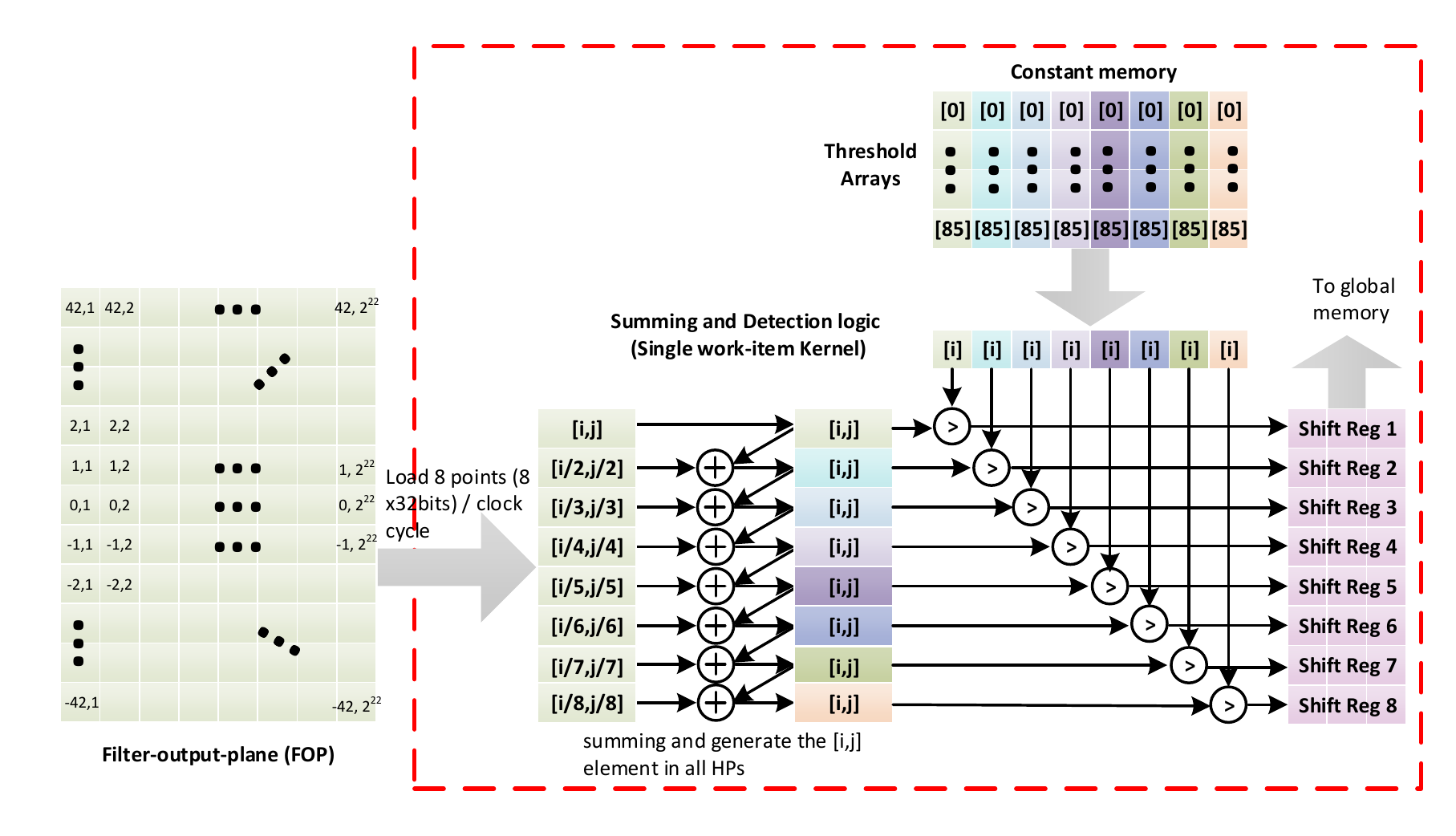}
\par\end{centering}
\caption{\label{fig:of-g}Architecture of the Na\text{\"i}ve \textsc{MultipleHP}
kernel (Single work-item) }
\end{figure}

The Na\text{\"i}ve\textsc{ MultipleHP} kernel calculates $N_{paral}$
points of all $N_{hp}$ harmonic planes with the same index, where
$N_{paral}$ is the parallelisation factor.   The architecture of
the Na\text{\"i}ve \textsc{MultipleHP} kernel is shown in Figure~\ref{fig:of-g},
where the operations in the red dot rectangle have to be parallelised
$N_{paral}$ times to process $N_{paral}$ points of all harmonic
planes. In OpenCL, this is implemented as a single work-item type,
and the \texttt{\#unroll pragma} $N_{paral}$ is added before the
main \texttt{for} loop in the kernel code.

The FOP is stored in global memory and $N_{hp}$ points ($\left(i,\,j\right)$,
$\left(\left\lfloor i/2\right\rfloor ,\,\left\lfloor j/2\right\rfloor \right)$,
...$,\left(\left\lfloor i/N_{hp}\right\rfloor ,\,\left\lfloor j/N_{hp}\right\rfloor \right)$)
are loaded in parallel to generate point $\left(i,\,j\right)$ of
all $N_{hp}$ harmonic planes. Then these $N_{hp}$ points are compared
with the corresponding thresholds, stored as constant memory. $N_{hp}$
independent arrays of size $N_{cand}$, one corresponding to each
harmonic plane, are employed to store the candidates. Both local memory
and shift register can be adopted to implement $N_{hp}$ arrays and
the performance difference is evaluated in Section~\ref{sec:Evaluation}.
After all $N_{hp}$ harmonic planes have been processed, the $N_{hp}$
candidate arrays are sent back to global memory.  Because the loading
accesses to the global memory are irregular, a high memory stall percentage
will impede the kernel from achieving a high performance. 

\subsubsection{ \textsc{MultipleHP-H}}

The \textsc{MultipleHP}\textbf{\textsc{-}}\textsc{H} kernel builds
on the Na\text{\"i}ve\textsc{ MultipleHP} kernel, which is a single
work-item kernel. \textsc{MultipleHP}\textbf{\textsc{-}}\textsc{H}
is however split into two parts, preloading and computing. The $N_{MultipleHP-H-preld}$
preloaded points that can be seen as constant cache memory are loaded
into a FIFO at runtime. In processing one FOP, there is no overlap
between the prefetching and computing parts. The available local memory
of the FPGA and the number of high touching frequency points affects
the performance of the \textsc{MultipleHP}\textbf{\textsc{-}}\textsc{H}
kernel. If the FOP size is comparable to the available local memory,
most of the points with high touching frequency can be loaded and
then most of the global memory accesses can be reduced. However, if
the number of high touching frequency points is significantly larger
than the local memory size, it is impossible for the device to hold
most of these important points. Besides these, the large proportion
of the used on-chip memory might lead to the decrease of kernel frequency.
In this case, it is necessary to search for the suitable $N_{MultipleHP-H-preld}$
for the target FPGA by testing a range of preloading data sizes. The
relationship between the $N_{MultipleHP-H-preld}$ and the kernel
performance is investigated in Section~\ref{sec:Evaluation}.

\subsubsection{\textsc{MultipleHP}-N}

The \textsc{MultipleHP}-N method is a memory accesses saving method,
as discussed in Section~\ref{subsec:Loading-Necessary-Points}. It
decomposes the overall task into a number of work-groups, and the
task for each work-group is to process $N_{MultipleHP-N-col}$ columns
of all harmonic planes. The NDRange kernel type is employed and the
preloading part of a work-group overlaps with the computing part of
the previous work-group. For the NDRange kernel, different work-groups
do not share local memory and it is inefficient to save candidates
in global memory during processing. The hybrid kernel type that contains
both single work-item type and NDRange type is employed to implement
the preloading necessary points kernel~(\textsc{MultipleHP}\textbf{\textsc{-}}N). 

The relationship between the work group size of the NDRange kernel
and the execution latency is studied next. The task of each work-group
is to generate $N_{MultipleHP-N-col}$ columns of all harmonic planes,
which contains $N_{MultipleHP-N-col}N_{chan}$ points. For each work-group,
$N_{hp}N_{MultipleHP-N-col}N_{chan}$ points are stored in local memory
using the OpenCL barrier technique (\texttt{\small{}barrier(CLK\_LOCAL\_MEM\_FENCE)}).
A number of points in these $N_{hp}N_{MultipleHP-N-col}N_{chan}$
points are from the same index in the FOP and they only need to be
loaded once. 

The NDRange harmonic plane calculation kernel is connected with the
single work-item candidate detection kernel through OpenCL channels,
which is a FIFO buffer in essence. The OpenCL channel is an effective
approach to transfer data between different kernels without touching
global memory. The candidate detection part is the same as that of
Na\text{\"i}ve \textsc{MultipleHP} kernel and \textsc{MultipleHP}-H
kernel.

\subsubsection{ \textsc{\label{subsec:-MultipleHP-R}MultipleHP-R}}

The \textsc{MultipleHP}\textbf{\textsc{-}}\textsc{R} kernel is based
on the \textsc{MultipleHP}\textbf{\textsc{-}}\textsc{N} kernel and
the main difference is the order of the data for each work-group.
After reordering, the points needed for a work-group are from consecutive
addresses. 

The total amount of needed data for a work-group ($N_{total/wg}$)
is the product of average needed data per column times the number
of columns per work-group ($N_{MultipleHP-R-col}$) (see also Figure~\ref{fig:Relationship-between-columns/wor-1}).
To achieve stream mode in global memory access, the number of loaded
points per clock cycle ($N_{lpoints/cc}$) has to be an integer constant,
which makes the product of $N_{lpoints/cc}$ and work-group size ($S_{workgroup}$)
usually larger than $N_{total/wg}$ and never less,
\[
N_{total/wg}\leq N_{lpoints/cc}S_{workgroup}.
\]
 In case of difference, the input array for each work-group has to
be padded with dummy values at the end. The relationship between $N_{lpoints/cc}$
and $N_{MultipleHP-R-col}$ is shown in Table~\ref{tab:loaded_n_G_O},
where $N_{points/wi}$ is the executed points of all harmonic planes
per work-item. The value in the bracket ($\times*$) represents the
ratio of total loaded points over the FOP size:
\[
\frac{N_{lpoints/cc}S_{workgroup}N_{workgroup}}{N_{chan}N_{temp}},
\]
 where $N_{workgroup}$ is the total number of work-groups. We use
\textsc{MultipleHP}\textbf{\textsc{-}}\textsc{R}-$(N_{MultipleHP-R-col},N_{points/wi})$
to represent kernel \textsc{MultipleHP}\textbf{\textsc{-}}\textsc{R}
with the specified settings. The larger $N_{MultipleHP-R-col}$ and
$N_{points/wi}$, the less data needs to be loaded from global memory.
Because of physical limitation, if the needed bandwidth of loading
$N_{lpoints/cc}$ points exceeds the total device off-chip memory
bandwidth, the performance will not increase and the kernel was not
implemented. 

\begin{table}
\caption{\label{tab:loaded_n_G_O}Number of loaded points per clock cycle $N_{lpoints/cc}$
of different $N_{points/wi}$ and $N_{MultipleHP-R-col}$ combinations
for general and optimised \textsc{MultipleHP}-R (number in ($\times*$)
shows total loaded points in relation to FOP size)}
\begin{centering}
{\footnotesize{}}
\par\end{centering}{\footnotesize \par}
\begin{centering}
{\footnotesize{}}%
\begin{tabular}{|c|c|c|c|c|c|}
\hline 
{\scriptsize{}$N_{points/wi}$} & \multirow{2}{*}{{\footnotesize{}Opt.}} & \multirow{2}{*}{{\footnotesize{}$\times1$}} & \multirow{2}{*}{{\footnotesize{}$\times2$}} & \multirow{2}{*}{{\footnotesize{}$\times4$}} & \multirow{2}{*}{{\footnotesize{}$\times8$}}\tabularnewline
\cline{1-1} 
{\scriptsize{}Columns} &  &  &  &  & \tabularnewline
\hline 
\hline 
\multirow{2}{*}{{\footnotesize{}1}} & {\footnotesize{}$\times$} & {\footnotesize{}3 ($\times3$)} & {\footnotesize{}6 ($\times3$)} & {\footnotesize{}12 ($\times3$)} & {\footnotesize{}23 ($\times2.9$)}\tabularnewline
 & {\footnotesize{}$\surd$} & {\footnotesize{}4 ($\times4$)} & {\footnotesize{}8 ($\times4$)} & {\footnotesize{}16 ($\times4$)} & {\footnotesize{}32 ($\times4$)}\tabularnewline
\hline 
\multirow{2}{*}{{\footnotesize{}4}} & {\footnotesize{}$\times$} & {\footnotesize{}2 ($\times2$)} & {\footnotesize{}4 ($\times2$)} & {\footnotesize{}8 ($\times2$)} & {\footnotesize{}15 ($\times1.9$)}\tabularnewline
 & {\footnotesize{}$\surd$} & {\footnotesize{}2 ($\times2$)} & {\footnotesize{}4 ($\times2$)} & {\footnotesize{}8 ($\times2$)} & {\footnotesize{}16 ($\times2$)}\tabularnewline
\hline 
\multirow{2}{*}{{\footnotesize{}16}} & {\footnotesize{}$\times$} & {\footnotesize{}2 ($\times2$)} & {\footnotesize{}4 ($\times2$)} & {\footnotesize{}7 ($\times1.8$)} & {\footnotesize{}13 ($\times1.6$)}\tabularnewline
 & {\footnotesize{}$\surd$} & {\footnotesize{}2 ($\times2$)} & {\footnotesize{}4 ($\times2$)} & {\footnotesize{}8 ($\times2$)} & {\footnotesize{}16 ($\times2$)}\tabularnewline
\hline 
\multirow{2}{*}{{\footnotesize{}64}} & {\footnotesize{}$\times$} & {\footnotesize{}2 ($\times2$)} & {\footnotesize{}4 ($\times2$)} & {\footnotesize{}7 ($\times1.8$)} & {\footnotesize{}13 ($\times1.6$)}\tabularnewline
 & {\footnotesize{}$\surd$} & {\footnotesize{}2 ($\times2$)} & {\footnotesize{}4 ($\times2$)} & {\footnotesize{}8 ($\times2$)} & {\footnotesize{}16 ($\times2$)}\tabularnewline
\hline 
\end{tabular}
\par\end{centering}{\footnotesize \par}
\end{table}

\begin{figure}
\begin{centering}
\includegraphics[bb=3bp 15bp 540bp 215bp,clip,scale=0.45]{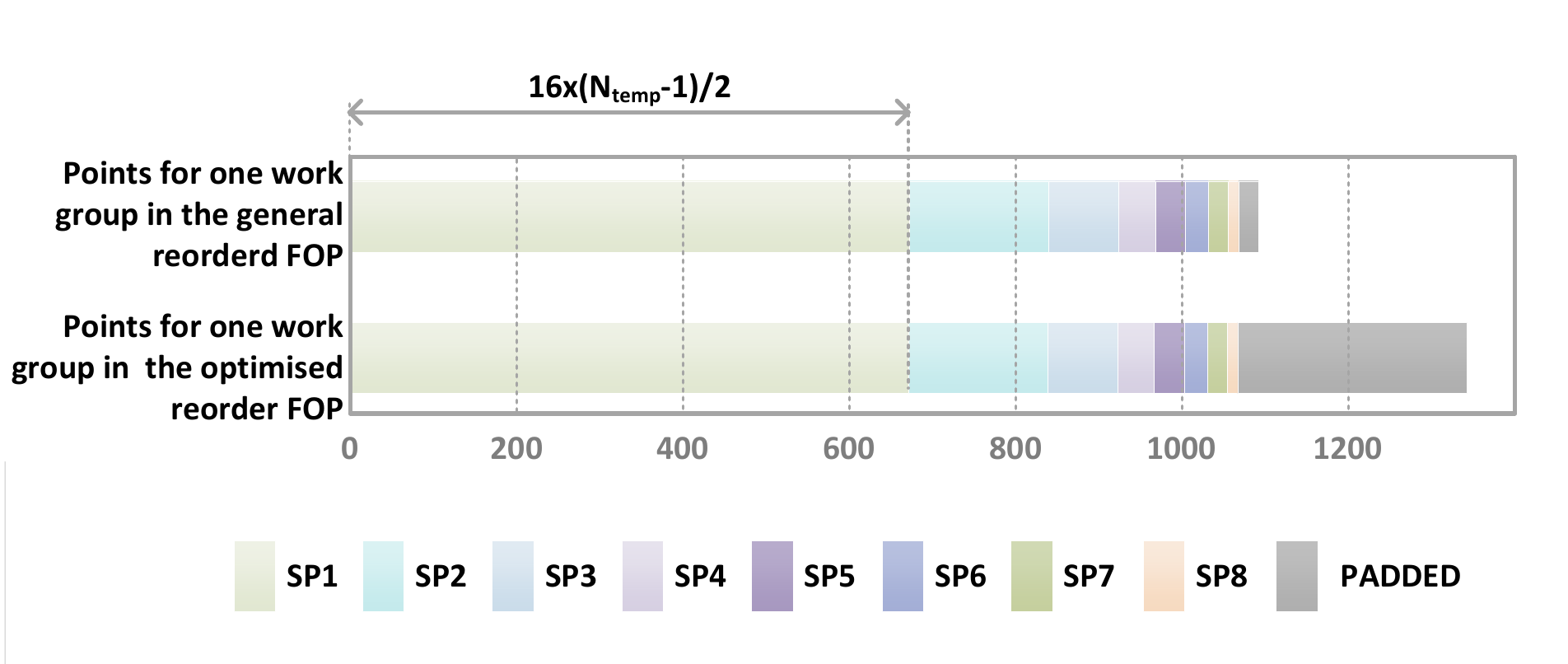}
\par\end{centering}
\caption{\label{fig:The-reordered-input-1}Needed points for one work group
of \textsc{MultipleHP}-R, the input array is reordered without optimising
$N_{points/wi}$ (top) and the optimised input array when $N_{points/wi}$
is a power of 2 (bottom)}
\end{figure}

It is clear that $N_{lpoints/cc}$, $N_{MultipleHP-R-col}$, and $N_{points/wi}$
are the three main parameters for kernel \textsc{MultipleHP}-R and
they have to be balanced to achieve good performance. Using the AOCL
compiler, it becomes apparent that using the number that is powers
of 2 for $N_{lpoints/cc}$ results in more efficient implementations
than other numbers. Hence, to make the value of $N_{lpoints/cc}$
equal a power of 2, more data might need to be loaded for each work
group. Take the kernel \textsc{MultipleHP}-R-$(8,8)$ for example,
the value of $N_{lpoints/cc}$ is 13, it has to be increased to the
nearest power of 2, which is 16. Since the number of loaded data per
work-group is $N_{lpoints/cc}S_{workgroup}$, the increase of $N_{lpoints/cc}$
leads to the increase of loading operations (as can be seen in the
example in Figure~\ref{fig:The-reordered-input-1}). The optimised
$N_{lpoints/cc}$, where $N_{lpoints/cc}$ is the lowest power of
2 greater or equal to the corresponding $N_{lpoints/cc}$ of values
without optimisation in Table~\ref{tab:loaded_n_G_O}. When $N_{MultipleHP-R-col}\geq4$,
the total loaded data is twice the FOP size (value in the bracket).

Take the $N_{MultipleHP-R-col}=16$ and half FOP as an example, the
input array needed for one work-group is depicted in Figure~\ref{fig:The-reordered-input-1},
where SP$i$ represents the needed points to form the $i$th stretch
plane and 'PADDED' are the dummy points to be padded at the end of
each array. It can be seen that when $N_{points/wi}$ is optimised
to a power of 2, more points need to be loaded during processing.

For the hybrid kernels (combining NDRange and single work-item kernels)
\textsc{MultipleHP}-H, \textsc{MultipleHP}-N, and \textsc{MultipleHP}-R,
adding attributes \texttt{\small{}num\_simd\_work\_items($N_{paral}$)}
or \texttt{\small{}num\_compute\_units($N_{paral}$) }can only parallelise
the NDRange part but not the single work-item part. To vectorize the
hybrid kernel and make it execute in a single instruction multiple
data (SIMD) fashion~\cite{altera2016openclpra}, it has to be parallelised
manually in the kernel code.

\subsection{Comparison}

The main challenge for FPGA devices in efficiently implementing the
harmonic-summing module is the global memory bandwidth and the number
of global memory accesses. In this section, we analyze the GMA of
the kernel discussed above and the GMB is evaluated in Section~\ref{sec:Evaluation}. 

For the \textsc{SingleHP} method, to process $N_{hp}$ harmonic planes,
the minimum number of FOP point accesses is
\[
\sum_{i=1}^{N_{hp}}\left\lceil \frac{N_{chan}}{i}\right\rceil \left\lceil \frac{N_{temp}-1}{i}\right\rceil .
\]
 Except for calculating $HP_{1}$, the loaded points from the FOP
have to be summed with the points of the $HP_{k}$ harmonic plane
and then the generated points of the next harmonic planes $HP_{k+1}$
are stored. The numbers of load and store operations of this part
are both $(N_{hp}-1)N_{temp}N_{chan}$ . Hence, the minimum amount
of global memory accesses for \textsc{SingleHP} method $GMA_{SingleHP-min}$is
the sum of these accesses: 
\begin{align*}
GMA_{SingleHP-min} & =\sum_{i=1}^{N_{hp}}\left\lceil \frac{N_{chan}}{i}\right\rceil \left\lceil \frac{N_{temp}}{i}\right\rceil \\
 & +2(N_{hp}-1)N_{temp}N_{chan}.
\end{align*}

In the\textsc{ MultipleHP} methods, only the FOP and candidates need
to be stored in global memory and the number of store operations
to global memory for all \textsc{MultipleHP} kernels is $0$. For
the Na\text{\"i}ve \textsc{MultipleHP} kernel, at most $N_{hp}$
points from FOP need to be loaded to calculate one point of the same
index in all $N_{hp}$ harmonic planes. In this case, the maximum
number of memory accesses is $GMA_{Na\text{{\"i}}ve-MultipleHP-max}=N_{hp}N_{temp}N_{chan}$. 

For the preloading high touching frequency method \textsc{MultipleHP}-H,
the GMA depends on $N_{MultipleHP-H-preld}$. The maximum amount of
memory accesses is $GMA_{MultipleHP-H-max}=N_{hp}N_{temp}N_{chan}-N_{MultipleHP-H-preld}$
and the minimum amount is to store the whole FOP in the local memory,
which means $GMA_{MultipleHP-H-min}=N_{temp}N_{chan}$. For the preloading
necessary points method \textsc{MultipleHP-N} and the reorder FOP
method \textsc{MultipleHP}-R, the $GMA_{MultipleHP-N}$ and $GMA_{MultipleHP-R}$
are both multiple of the FOP size $N_{temp}N_{chan}$. Table~ \ref{tab:Operation-between-FPGA}
summarizes the GMA of the different kernels. $C_{0}$, $C_{1}$ and
$C_{2}$ are all constants and no less than 1. The range of $C_{0}$
is $\left[1,\,N_{hp}N_{temp}N_{chan}/N_{MultipleHP-H-preld}\right]$.
For $C_{1}$ and $C_{2}$, $C_{1}\leq C_{2}$ and, in Table~\ref{tab:loaded_n_G_O},
$C_{2}=2$ when $N_{MultipleHP-R-col}\geq4$.

\begin{table}
\caption{\label{tab:Operation-between-FPGA}Number of accesses to and from
global memory }
\centering{}{\footnotesize{}}%
\begin{tabular}{|c|c|c|}
\hline 
{\footnotesize{}Kernels} & {\footnotesize{}GMA (Store)} & {\footnotesize{}GMA (Load) }\tabularnewline
\hline 
\hline 
\multirow{2}{*}{\textsc{\footnotesize{}SingleHP}{\footnotesize{} }} & {\footnotesize{}$(N_{hp}-1)\times$} & {\footnotesize{}$N_{hp}N_{temp}N_{chan}+$ }\tabularnewline
 & {\footnotesize{}$N_{temp}N_{chan}$} & {\footnotesize{}$\sum_{i=2}^{N_{hp}}\left\lceil \frac{N_{chan}}{i}\right\rceil \left\lceil \frac{N_{temp}}{i}\right\rceil $}\tabularnewline
\hline 
{\footnotesize{}Na\text{\"i}ve}\textsc{\footnotesize{} } & \multirow{2}{*}{{\footnotesize{}0}} & \multirow{2}{*}{{\footnotesize{}$N_{hp}N_{temp}N_{chan}$}}\tabularnewline
\textsc{\footnotesize{}MultipleHP} &  & \tabularnewline
\hline 
\multirow{2}{*}{\textsc{\footnotesize{}MultipleHP}{\footnotesize{}-H}} & \multirow{2}{*}{{\footnotesize{}0}} & {\footnotesize{}$N_{hp}N_{temp}N_{chan}-$}\tabularnewline
 &  & {\footnotesize{}$C_{0}N_{MultipleHP-H-preld}$}\tabularnewline
\hline 
\textsc{\footnotesize{}MultipleHP}{\footnotesize{}-N} & {\footnotesize{}0} & {\footnotesize{}$C_{1}N_{temp}N_{chan}$}\tabularnewline
\hline 
\textsc{\footnotesize{}MultipleHP}{\footnotesize{}-R} & {\footnotesize{}0} & {\footnotesize{}$C_{2}N_{temp}N_{chan}$}\tabularnewline
\hline 
\end{tabular}{\footnotesize \par}
\end{table}

The load accesses of the \textsc{SingleHP }kernel $GMA_{SingleHP-min}$
is larger than the overall point accesses (store+load) of each \textsc{MultipleHP}
method based kernel. This is the major advantage of the\textsc{ MultipleHP}
method over the \textsc{SingleHP} method.

\section{\label{sec:Evaluation}Experimental Evaluation}

To experimentally evaluate the harmonic-summing module, the straightforward
\textsc{SingleHP} method and the proposed\textsc{ MultipleHP}-based
methods are evaluated in this section. The FPGA-based harmonic-summing
kernels are assessed according to their resource usage, execution
latency, and energy dissipation. Additionally, we compare those results
to latency and energy dissipation of the kernels implemented on GPU
and multicore CPUs. 

\subsection{Experimental Setup}

Four different devices are employed to evaluate the performance of
the proposed designs on CPU, GPU, and FPGAs. Two types of Intel FPGAs
(Stratix V, referred to as $\mathbf{S5}$, and Arria 10, referred
to as $\mathbf{A10}$) are compared with one mid-range AMD R7 GPU,
referred to as $\mathbf{R7}$, and a general Intel $i7$ CPU, referred
to as $\mathbf{I7}$. The specifications of these platforms are given
in Table~\ref{tab:Details-of-FPGA}. The FPGA and GPU cards are connected
to the host processor through the PCIe bus.

\begin{table*}
\caption{\label{tab:Details-of-FPGA}Specifications of CPU, GPU and FPGA platforms}
\centering{}{\small{}}%
\begin{tabular}{|c|c|c|c|c|}
\hline 
{\footnotesize{}Device} & {\footnotesize{}Terasic DE5-Net (}\textbf{\footnotesize{}S5}{\footnotesize{})} & {\footnotesize{}Nallatech 385A (}\textbf{\footnotesize{}A10}{\footnotesize{})} & {\footnotesize{}Sapphire Nitro R7 370 (}\textbf{\footnotesize{}R7}{\footnotesize{})} & {\footnotesize{}Intel CPU Host (}\textbf{\footnotesize{}I7}{\footnotesize{})}\tabularnewline
\hline 
\hline 
{\footnotesize{}Hardware} & {\footnotesize{}Intel Stratix V 5SGXA7} & {\footnotesize{}Intel Arria 10 GX1150} & {\footnotesize{}AMD Radeon R7 370} & {\footnotesize{}Intel Core i7-6700K}\tabularnewline
\hline 
{\footnotesize{}Technology} & {\footnotesize{}$28nm$} & {\footnotesize{}$20nm$} & {\footnotesize{}$28nm$} & {\footnotesize{}$14nm$}\tabularnewline
\hline 
\multirow{2}{*}{{\footnotesize{}Compute resource}} & {\footnotesize{}622,000 LEs} & {\footnotesize{}1,506,000 LEs} & {\footnotesize{}1,024 Stream Processors} & {\footnotesize{}8 Processors}\tabularnewline
 & {\footnotesize{}256 DSP blocks} & {\footnotesize{}1,518 DSP blocks} & {\footnotesize{}(16 Compute Units)} & {\footnotesize{}(4 Cores)}\tabularnewline
\hline 
{\footnotesize{}On-chip memory size} & {\footnotesize{}50$Mb$} & {\footnotesize{}53$Mb$} & {\footnotesize{}\textemdash{}} & {\footnotesize{}64$Mb$}\tabularnewline
\hline 
{\footnotesize{}Off-chip memory size} & {\footnotesize{}2 x $2GB$ DDR3} & {\footnotesize{}2 x $4GB$ DDR3} & {\footnotesize{}$4GB$ GDDR5} & {\footnotesize{}$64GB$ DDR4}\tabularnewline
\hline 
{\footnotesize{}Memory interface width} & {\footnotesize{}2 x 64-bit} & {\footnotesize{}2 x 72-bit} & {\footnotesize{}256-bit} & {\footnotesize{}\textemdash{}}\tabularnewline
\hline 
{\footnotesize{}Max clock frequency} & {\footnotesize{}600$MHz$} & {\footnotesize{}1.5$GHz$} & {\footnotesize{}985$MHz$} & {\footnotesize{}4.2$GHz$}\tabularnewline
\hline 
{\footnotesize{}Max power consumption} & {\footnotesize{}\textemdash{}} & {\footnotesize{}75W} & {\footnotesize{}150W} & {\footnotesize{}\textemdash{}}\tabularnewline
\hline 
\end{tabular}{\small \par}
\end{table*}

All FPGA-targeting OpenCL kernels are compiled using AOCL version
16.0.0.222 and GPU-targeting kernels are compiled using AMD APP SDK
version 3.0. For the CPU platform, the C code, which is based on the
same kernel code, is compiled using GCC, using OpenMP for parallelisation. 

Since the top half (from row 1 to $\frac{N_{temp}-1}{2}$) and the
bottom half (from row $\frac{1-N_{temp}}{2}$ to -1) are independent
for the harmonic-summing module, we investigate the performance, in
terms of the execution latency and energy dissipation, of half of
the FOP as specified in Table~\ref{tab:Harmonic-summing-Module-Paramete},
which size is $42\times2^{21}$. Remember from Section~\ref{sec:Harmonic-summing-Module-and}
that the upper and lower half of the FOP can be processed independently
and the required processing is identical. The size of candidate list
is 200.

\subsection{\label{subsec:Resource-Usage}Resource Usage}

Because the harmonic-summing module is not a compute-intensive application,
the DSP block utilization of all implementations is less than 5\%.
We discuss the logic utilization, RAM blocks utilization, and kernel
frequency in this section.

\subsubsection{\textsc{SingleHP}}

\begin{figure}
\begin{centering}
\includegraphics[bb=0bp 15bp 270bp 240bp,clip,scale=0.45]{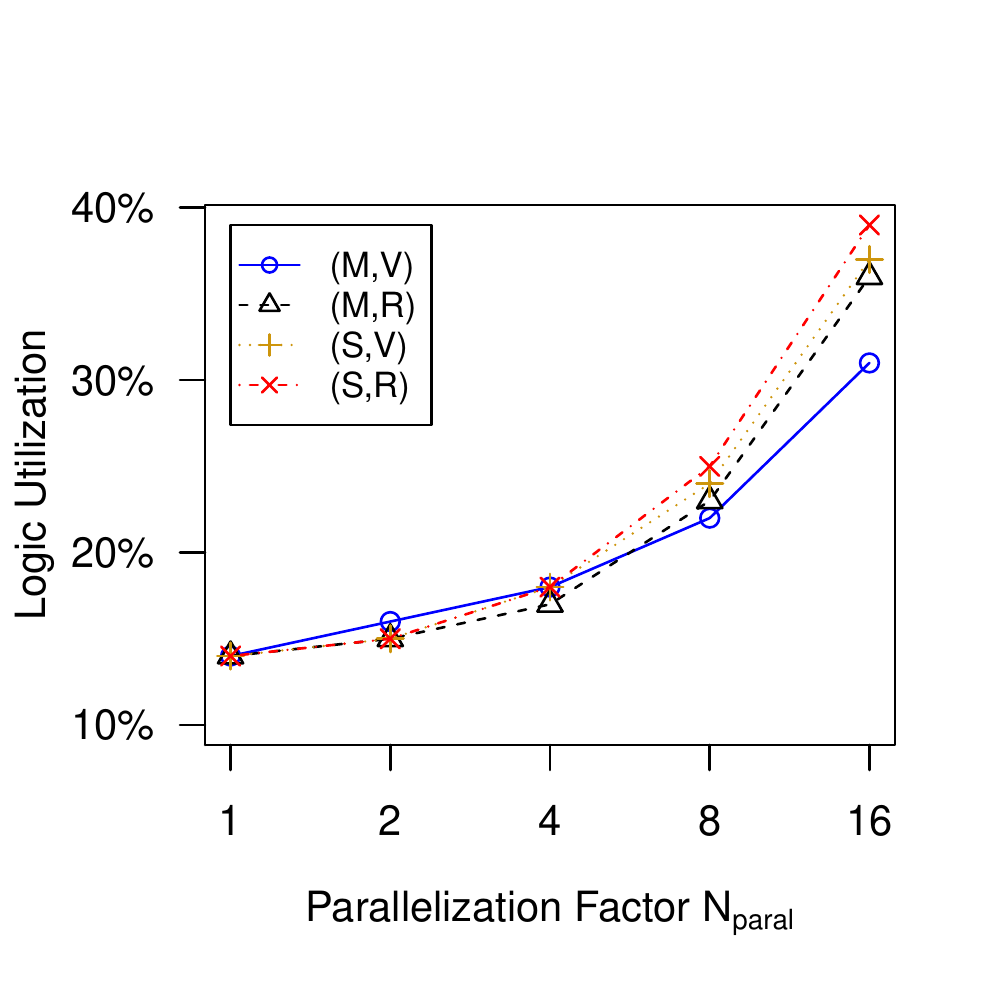}\includegraphics[bb=0bp 15bp 270bp 240bp,clip,scale=0.45]{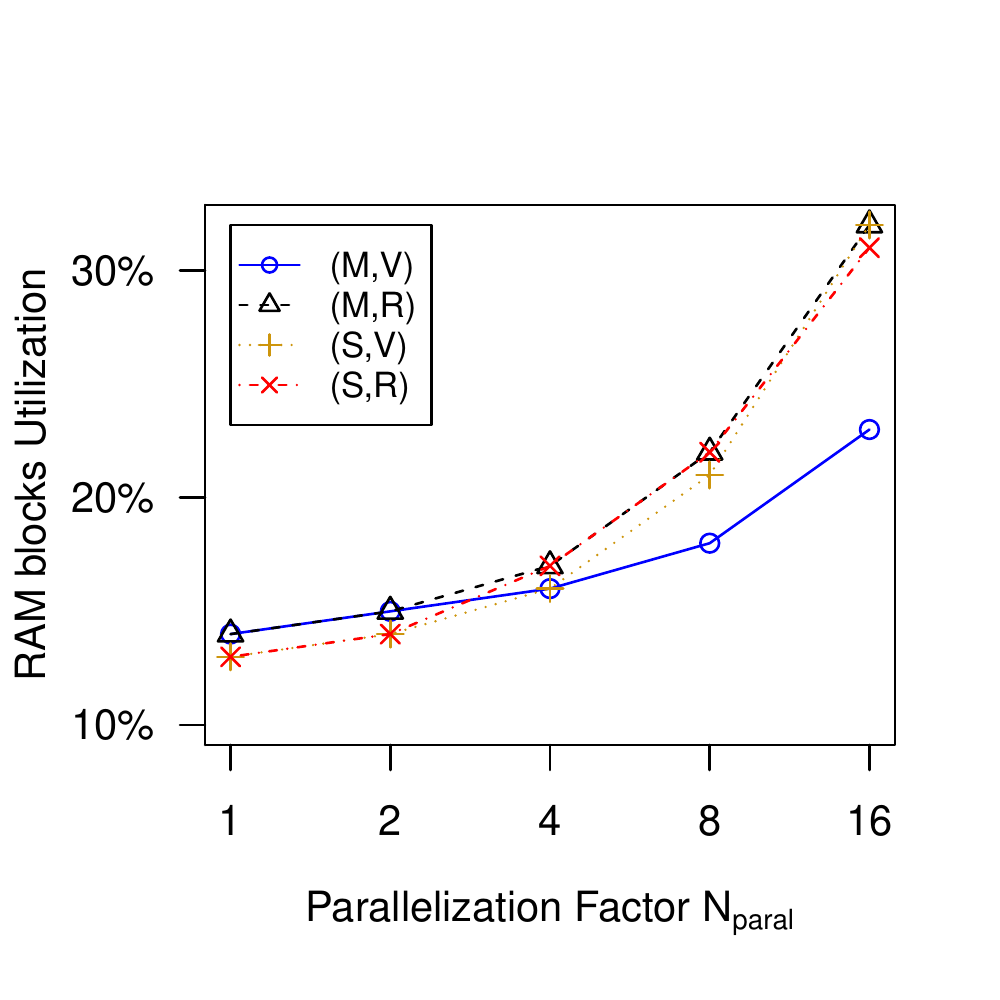}
\par\end{centering}
\begin{centering}
\par\end{centering}
\caption{\label{fig:singlehp-resource usage}Logic utilization and RAM block
usage of \textsc{SingleHP} kernels on $A10$}
\end{figure}

A series of \textsc{SingleHP} kernels with different parallelisation
factors $N_{paral}$ are evaluated. These kernels are employed to
generate eight harmonic planes of half FOP. All these kernels are
NDRange kernels and the work group sizes are set to 256. The usage
of logic cells and RAM blocks of these kernels are given in Figure~\ref{fig:singlehp-resource usage},
where '$S$' and '$M$' represent single launch and multiple launches,
and '$V$' and '$R$' represent kernel vectorization and replication.
The candidate detection part is included, and the local memory is
employed to store the candidate during processing. When $N_{paral}=1$,
it means the kernel is not parallelised and that vectorization and
replication are not employed. 

It can be seen that the usage of both resources increases as $N_{paral}$
increases. These trends are similar to those observed for execution
on $S5$. The kernel frequency drops as the resource usage increases
across all kernels. Take the kernel \textsc{SingleHP-$(M,V)$} on
$A10$ as an example, its frequency decreases from $266.9MHz$ at
$N_{paral}=1$ to $236.8MHz$ at $N_{paral}=16$. 

\subsubsection{\textsc{MultipleHP}}

In terms of the \textsc{MultipleHP} designs, Na\text{\"i}ve\textsc{
MultipleHP}, \textsc{MultipleHP}-H, \textsc{MultipleHP}-H, and \textsc{MultipleHP}-R
(Section~\ref{sec:OpenCL-based-Architecture}) are evaluated.

\paragraph{Na\text{\"i}ve\textsc{ MultipleHP }and\textsc{ MultipleHP-H}}

\begin{figure}
\begin{centering}
\includegraphics[bb=0bp 15bp 504bp 240bp,clip,scale=0.5]{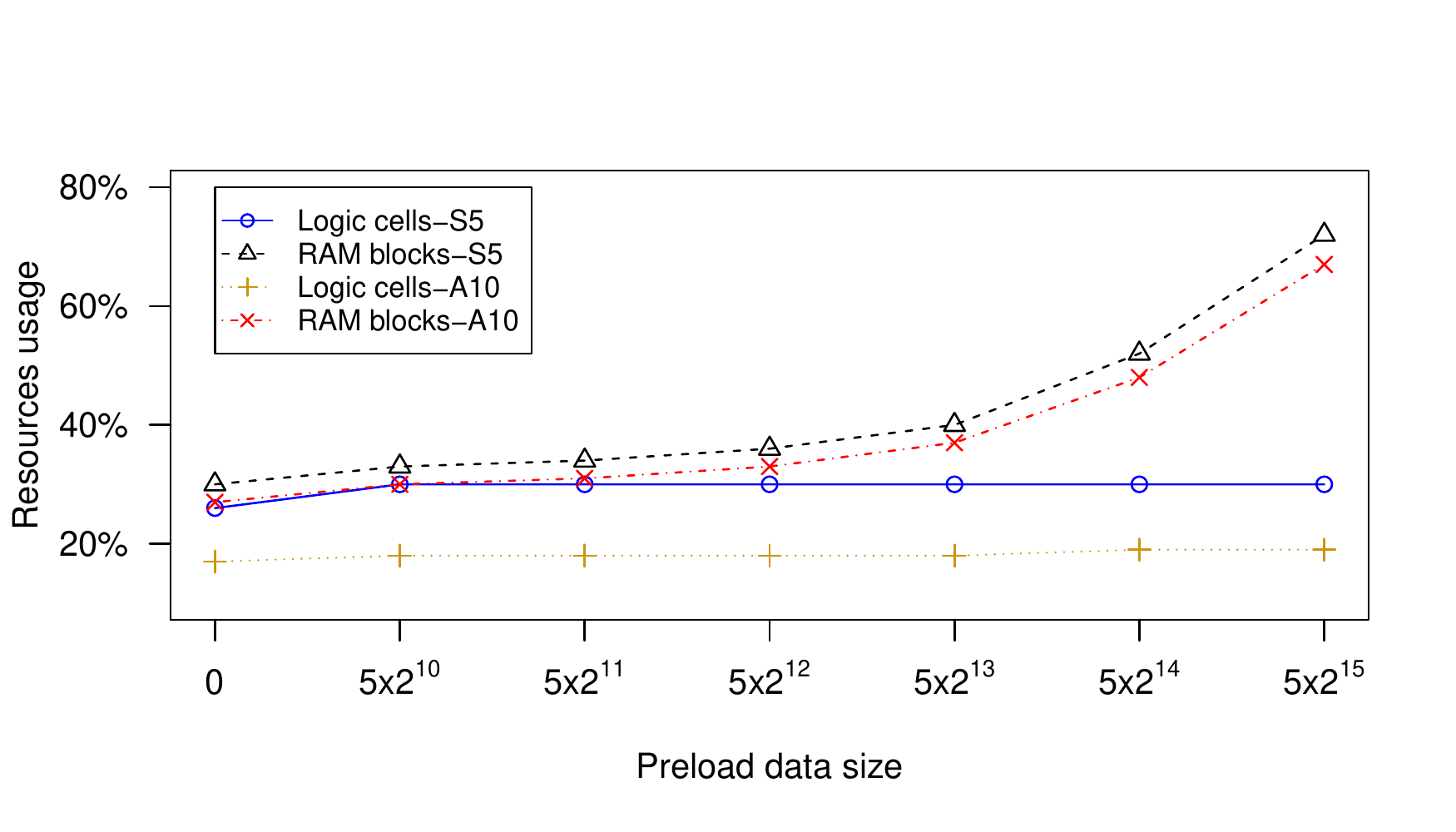}
\par\end{centering}
\caption{\label{fig:Resource-of-MultipleHP-H}Resource usage of \textsc{MultipleHP-H
}with different $N_{MultipleHP-H-preld}$ on $S5$ and $A10$}
\end{figure}

The \textsc{MultipleHP}-H is based on the Na\text{\"i}ve\textsc{
MultipleHP-H, }and the main difference is that it preloads a block
of data before calculating. The resource usages of these kernels is
plotted over the preloaded data size in Figure~\ref{fig:Resource-of-MultipleHP-H}.
The value points for $N_{MultipleHP-H-preld}=0$ correspond to Na\text{\"i}ve\textsc{
MultipleHP}. The logic utilization is not affected by the increase
of $N_{MultipleHP-H-preld}$, however, the RAM blocks utilization
increases. The kernel frequency is around $210MHz$ for $S5$ based
implementations and $220MHz$ for $A10$ based implementations.

\paragraph{\textsc{MultipleHP-N }and \textsc{MultipleHP-R}}

In contrast to \textsc{MultipleHP}-H, kernel \textsc{MultipleHP}-N
and \textsc{MultipleHP}-R do not depend heavily on local memory size.
\textsc{MultipleHP}-R is based on \textsc{MultipleHP}-N, however,
it does not need to load points from different locations.

For \textsc{MultipleHP}-N, different column numbers ($N_{MultipleHP-N-col}$)
are evaluated, and the results are listed in Table~\ref{tab:Resource-Multiple-n}.
As can be seen with increasing $N_{MultipleHP-N-col}$ both logic
cell and RAM block utilization increase. For most of the kernels,
the kernel frequency is decreased as $N_{MultipleHP-N-col}$ increases.

\begin{table}
\caption{\label{tab:Resource-Multiple-n}Resource usage and kernel frequency
of \textsc{MultipleHP}-N with candidate detection on $A10$}
\begin{centering}
{\footnotesize{}}
\par\end{centering}{\footnotesize \par}
\centering{}{\footnotesize{}}%
\begin{tabular}{|c|c|c|c|c|c|}
\hline 
{\footnotesize{}Columns} & {\footnotesize{}1} & {\footnotesize{}2} & {\footnotesize{}4} & {\footnotesize{}6} & {\footnotesize{}8}\tabularnewline
\hline 
\hline 
{\footnotesize{}Logic cells} & {\footnotesize{}17\%} & {\footnotesize{}25\%} & {\footnotesize{}25\%} & {\footnotesize{}29\%} & {\footnotesize{}30\%}\tabularnewline
\hline 
{\footnotesize{}RAM blocks} & {\footnotesize{}19\%} & {\footnotesize{}44\%} & {\footnotesize{}49\%} & {\footnotesize{}56\%} & {\footnotesize{}69\%}\tabularnewline
\hline 
{\footnotesize{}Frequency } & \multirow{2}{*}{{\footnotesize{}276.54}} & \multirow{2}{*}{{\footnotesize{}193.38}} & \multirow{2}{*}{{\footnotesize{}171.11}} & \multirow{2}{*}{{\footnotesize{}148.67}} & \multirow{2}{*}{{\footnotesize{}165.48}}\tabularnewline
{\footnotesize{}$(MHz)$} &  &  &  &  & \tabularnewline
\hline 
{\footnotesize{}Latency} & \multirow{2}{*}{{\footnotesize{}328.0}} & \multirow{2}{*}{{\footnotesize{}469.0}} & \multirow{2}{*}{{\footnotesize{}530.1}} & \multirow{2}{*}{{\footnotesize{}610.1}} & \multirow{2}{*}{{\footnotesize{}548.1}}\tabularnewline
{\footnotesize{}$(ms)$} &  &  &  &  & \tabularnewline
\hline 
\end{tabular}{\footnotesize \par}
\end{table}

Regarding \textsc{MultipleHP}-R, to arrange the data for each work
group into a consecutive address area, the half \textsc{FOP} is reordered
into a half \textsc{rFOP} (Section~\ref{subsec:MultipleHP}), in
the host program using \texttt{memcpy()}. The reordering latency on
the employed host is $87.8ms$ and the performance of two variants
of \textsc{MultipleHP}-R kernels (generating 16 and 64 columns of
all eight harmonic planes per work group) are evaluated, which is
shown in Table~\ref{tab:nc/wg=00003D16 64-1}. Four different points
per work-item values $N_{points/wi}$ (1, 2, 4, and 8) are tested
in this research. Since the values of $N_{lp/cc}$ for $N_{points/wi}=1$
and $N_{points/wi}=2$ are already powers of 2, so we focus on the
other two conditions ($N_{points/wi}=4$ and $N_{points/wi}=8$) and
the resource usage of the general and the optimised implementations
with these values are given in Table~\ref{tab:nc/wg=00003D16 64-1}.
For the optimised implementations, the values of $N_{lpoints/cc}$
are powers of 2 and this costs fewer logic cells than the general
implementations. Since more points are loaded per clock cycle, the
optimised implementations consume more RAM blocks. Besides these,
the kernel frequency of the optimised implementations is higher than
that of general implementations. 

Since $N_{MultipleHP-R-col}$, $N_{p/wi}$, and $N_{lp/cc}$ are three
main parameters that affect the performance of \textsc{MultipleHP}-R,
we investigate the trend of changing these parameters, but here without
candidate detection, hence the values in Table~\ref{tab:nc/wg=00003D16 64-1}
are only for the NDRange part. We do this because after combining
with the candidate detection, some of the \textsc{MultipleHP}-R kernels
such as \textsc{MultipleHP}-R-$(64,8)$ cannot be compiled because
of the limited resources, and we wanted to explore the influence of
the parameters in a good range. We employ the \textsc{MultipleHP}-R-$(16,4)$
kernel with candidate detection, which can be compiled on both $S5$
and $A10$, to compare with other methods. In the future, as FPGA
technology upgrades, the amount of on-chip logic cells and RAM blocks
increase. The values of $N_{MultipleHP-R-col}$ and $N_{points/wi}$
can be raised, and the execution latency is likely to be faster than
that achieved in Table~\ref{tab:nc/wg=00003D16 64-1}.

\begin{table*}
\caption{\label{tab:nc/wg=00003D16 64-1}Resource usage and execution latency
of\textsc{ MultipleHP}-R (NDRange part only) with ($N_{lp/cc}$ is
power of 2) and without optimising GMB on $A10$ (\textit{without
candidate detection})}
\begin{centering}
{\footnotesize{}}%
\begin{tabular}{|c|c|c|c|c|c|c|c|c|c|c|c|}
\hline 
\multicolumn{2}{|c|}{{\footnotesize{}$N_{MultipleHP-R-col}$}} & \multicolumn{5}{c|}{{\footnotesize{}$16$}} & \multicolumn{5}{c|}{{\footnotesize{}$64$}}\tabularnewline
\hline 
\hline 
\multirow{2}{*}{{\footnotesize{}$N_{p/wi}$}} & \multirow{2}{*}{{\footnotesize{}$N_{lp/cc}$}} & {\footnotesize{}Logic } & {\footnotesize{}RAM } & {\footnotesize{}Freq.} & {\footnotesize{}Latency} & {\footnotesize{}Occup.} & {\footnotesize{}Logic } & {\footnotesize{}RAM } & {\footnotesize{}Freq.} & {\footnotesize{}Latency} & {\footnotesize{}Occup.}\tabularnewline
 &  & {\footnotesize{}utilization} & {\footnotesize{}blocks} & {\footnotesize{}($MHz$)} & {\footnotesize{}$(ms)$} &  & {\footnotesize{}utilization} & {\footnotesize{}blocks} & {\footnotesize{}($MHz$)} & {\footnotesize{}$(ms)$} & \tabularnewline
\hline 
{\footnotesize{}1} & {\footnotesize{}2} & {\footnotesize{}14\%} & {\footnotesize{}12\%} & {\footnotesize{}269.2} & {\footnotesize{}350.8} & {\footnotesize{}93.4\%} & {\footnotesize{}15\%} & {\footnotesize{}12\%} & {\footnotesize{}266.7} & {\footnotesize{}336.2} & {\footnotesize{}98.3\%}\tabularnewline
\hline 
{\footnotesize{}2} & {\footnotesize{}4} & {\footnotesize{}15\%} & {\footnotesize{}13\%} & {\footnotesize{}286.5} & {\footnotesize{}176.7} & {\footnotesize{}87\%} & {\footnotesize{}16\%} & {\footnotesize{}12\%} & {\footnotesize{}252.8} & {\footnotesize{}180.9} & {\footnotesize{}96.6\%}\tabularnewline
\hline 
{\footnotesize{}4} & {\footnotesize{}7} & {\footnotesize{}22\%} & {\footnotesize{}14\%} & {\footnotesize{}196.3} & {\footnotesize{}189.1} & {\footnotesize{}59.4\%} & {\footnotesize{}22\% } & {\footnotesize{}15\%} & {\footnotesize{}161.9} & {\footnotesize{}248.1} & {\footnotesize{}55\%}\tabularnewline
{\footnotesize{}4} & {\footnotesize{}8} & {\footnotesize{}19\%} & {\footnotesize{}16\%} & {\footnotesize{}263.0} & {\footnotesize{}107.7} & {\footnotesize{}77.8\%} & {\footnotesize{}19\%} & {\footnotesize{}30\%} & {\footnotesize{}229.6} & {\footnotesize{}102.5} & {\footnotesize{}93.8\%}\tabularnewline
\hline 
{\footnotesize{}8} & {\footnotesize{}13} & {\footnotesize{}35\%} & {\footnotesize{}17\%} & {\footnotesize{}130.3} & {\footnotesize{}163.9} & {\footnotesize{}51.6\%} & {\footnotesize{}37\%} & {\footnotesize{}17\%} & {\footnotesize{}135.1} & {\footnotesize{}158.4} & {\footnotesize{}51.6\%}\tabularnewline
{\footnotesize{}8} & {\footnotesize{}16} & {\footnotesize{}28\%} & {\footnotesize{}29\%} & {\footnotesize{}168.1} & {\footnotesize{}93.1} & {\footnotesize{}70.5\%} & {\footnotesize{}29\%} & {\footnotesize{}48\%} & {\footnotesize{}171.4} & {\footnotesize{}71.7} & {\footnotesize{}89.7\%}\tabularnewline
\hline 
\end{tabular}
\par\end{centering}{\footnotesize \par}
\end{table*}

\subsubsection{}

\subsection{Latency Evaluation}

\subsubsection{\textsc{\label{subsec:SingleHP-and-MultipleHP}}Harmonic Plane Calculation
on FPGA}

To find the suitable design for a specific device, we evaluate the
overall execution latency of the harmonic-summing module, including
the harmonic plane calculation and the candidate detection. The points
of the 8th harmonic plane are compared with the result of a Matlab
implementation to verify the correctness of the harmonic plane calculation
in the different designs.

\paragraph{\textsc{SingleHP}}

\begin{figure}
\begin{centering}
\includegraphics[bb=0bp 15bp 270bp 240bp,clip,scale=0.45]{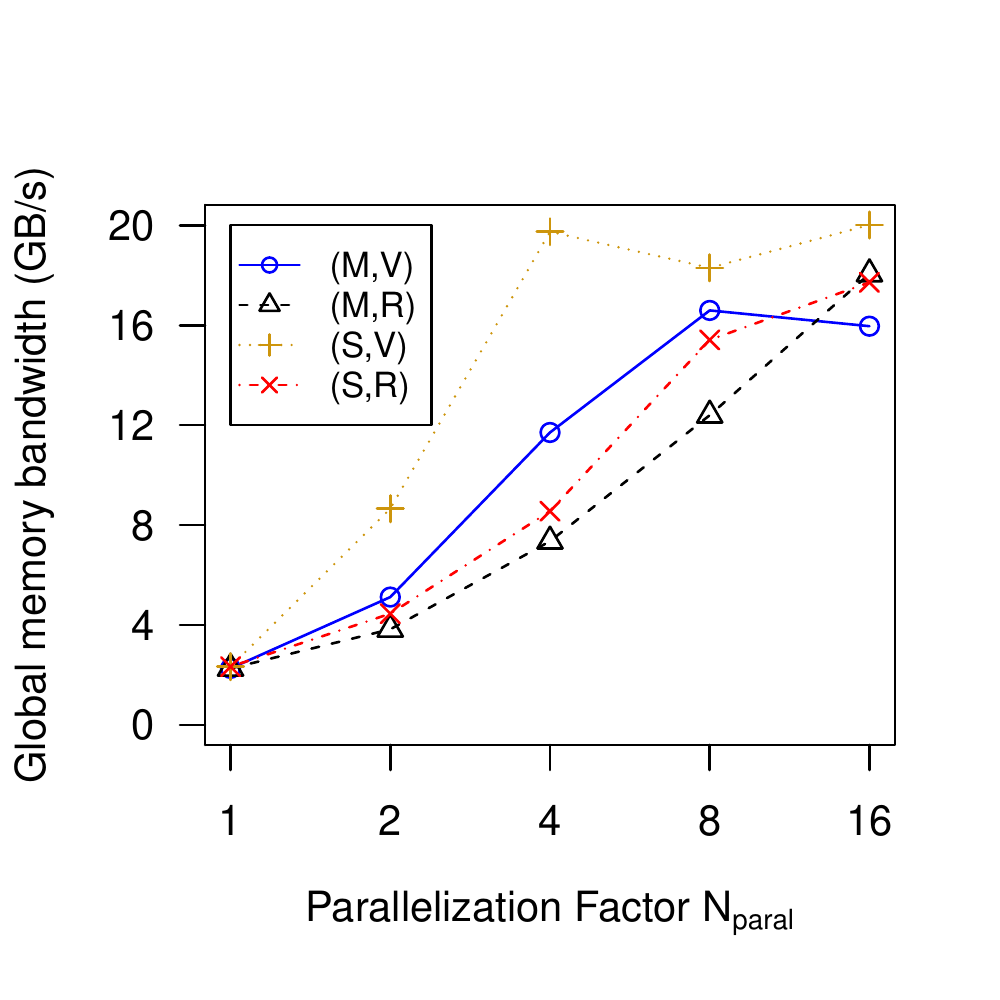}\includegraphics[bb=0bp 15bp 270bp 240bp,clip,scale=0.45]{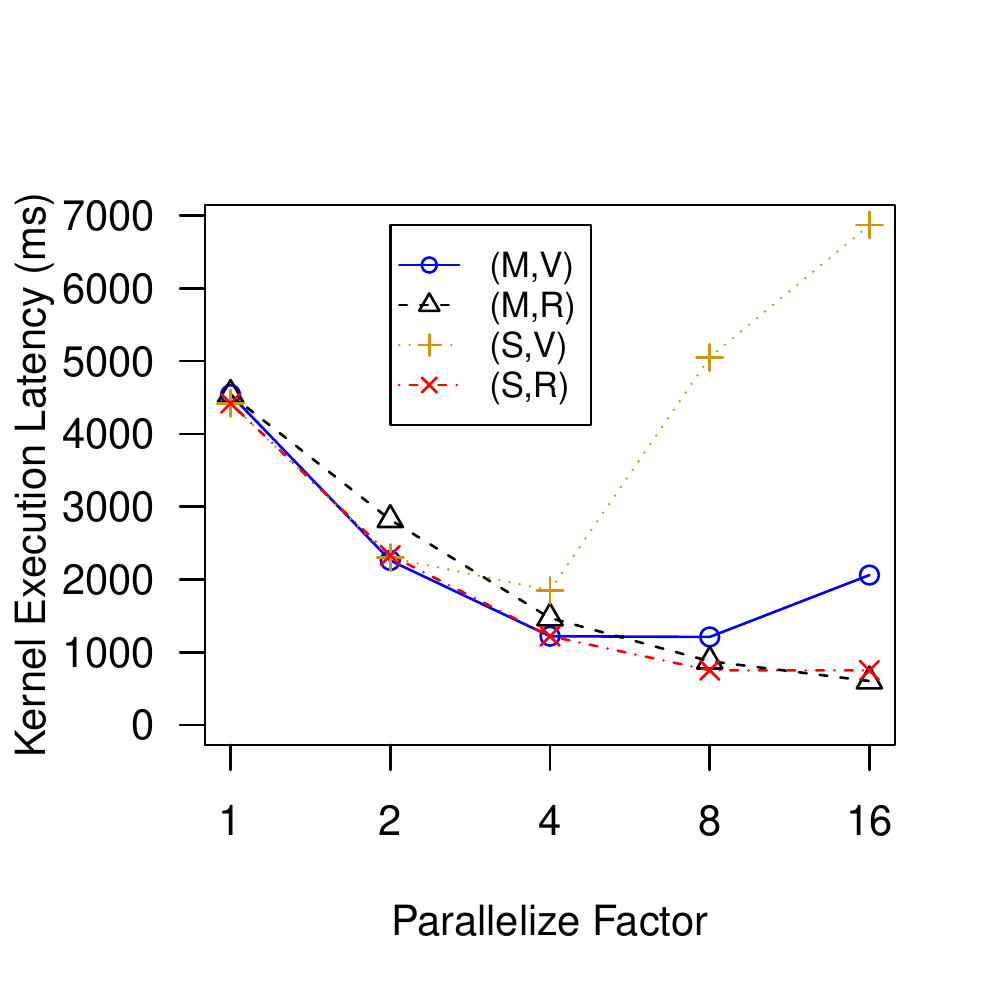}
\par\end{centering}
\caption{\label{fig:s-and-execution}GMBs and execution latency of \textsc{SingleHP}
on $A10$}
\end{figure}

The used GMBs and execution latencies of the \textsc{SingleHP} kernel
with various $N_{paral}$ in Section~\ref{subsec:Resource-Usage}
are shown in Figure~\ref{fig:s-and-execution}. As $N_{paral}$ increases
the GMBs of all \textsc{SingleHP} kernels increase, however, not all
execution latencies are decreased. 

For the two multiple launches ('$M$') kernels \textsc{SingleHP}-($M,V$)
and \textsc{SingleHP}-($M,R$), the launching overhead is hundreds
of times smaller than the kernel execution latency and hence negligible.
For the two single launch kernels \textsc{SingleHP}-($S,V$) and \textsc{SingleHP}-($S,R$),
the performance stops increasing when $N_{paral}$ is larger than
8. When $N_{paral}=8$, kernel \textsc{SingleHP}-($S,R$) performs
better than other kernels and the \textsc{SingleHP}-($M,R$) kernel
performs best when $N_{paral}=16$, which is about $7.5$ times faster
than \textsc{SingleHP}-($M,\,$) with $N_{paral}=1$. 

\paragraph{Na\text{\"i}ve\textsc{ MultipleHP}}

The execution latency of kernel Na\text{\"i}ve\textsc{ MultipleHP}
on $S5$ is over one second ($1,210ms$), however, the same kernel
achieves a better performance, which is less than $400ms$ on $A10$.
The main reason is the kernel frequency achieved on $A10$ is over
two times higher than that on $S5$. This might be caused by the board
support packages (BSPs) provided by different vendors.

\paragraph{\textsc{MultipleHP-H}}

\begin{figure}
\begin{centering}
\includegraphics[bb=0bp 15bp 504bp 240bp,clip,scale=0.5]{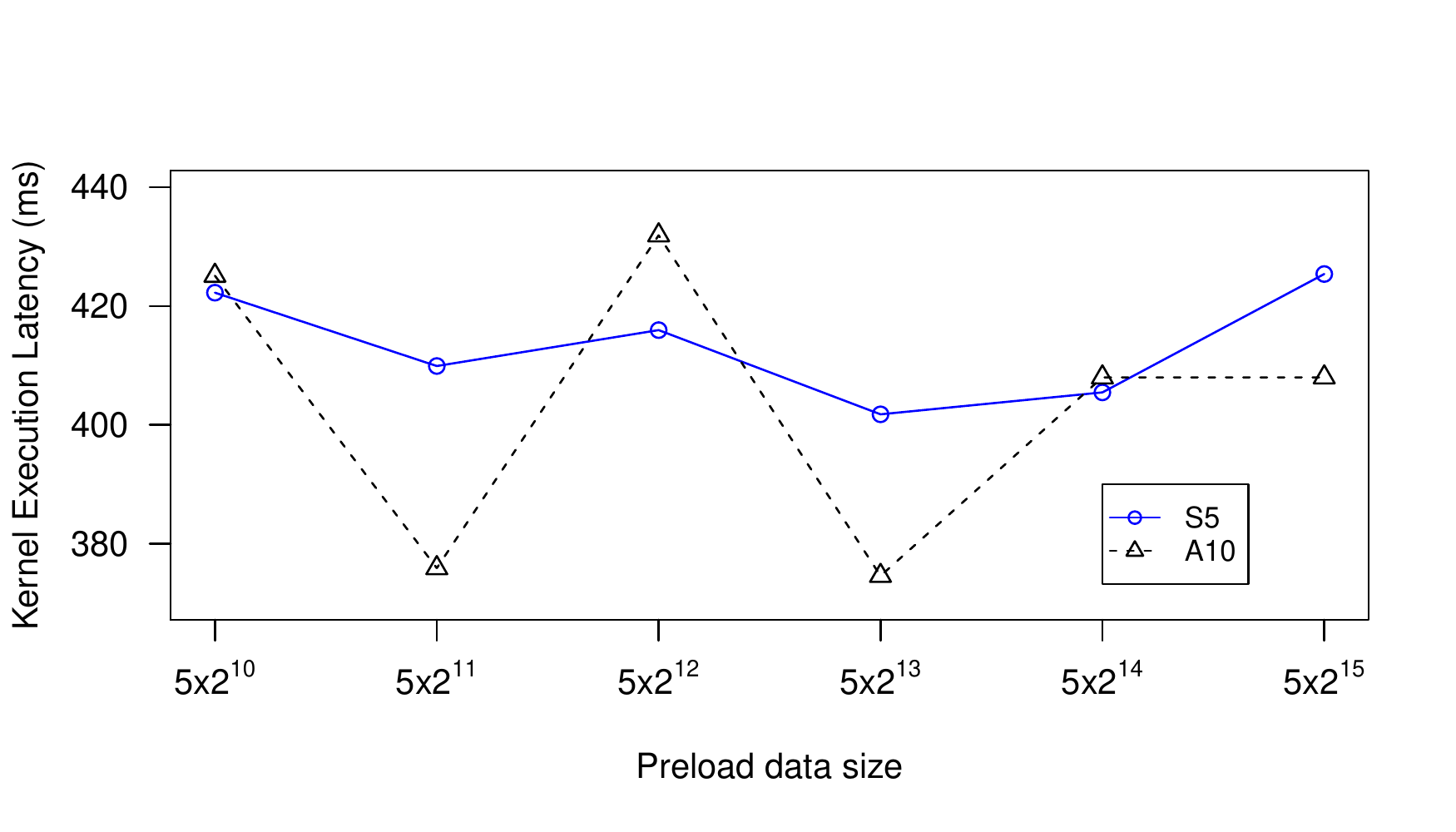}
\par\end{centering}
\caption{\label{fig:Execution-latencies-of}Execution latencies of the \textsc{MultipleHP}-H
kernels with different sizes of preloaded points}
\end{figure}

The relationship between the number of preloaded data points $N_{MultipleHP-H-preld}$
and the execution latency of \textsc{MultipleHP}-H is investigated
on both $S5$ and $A10$. The half FOP is transposed and then processed
row by row (each row has $\frac{N_{temp}-1}{2}$ points). The execution
latencies of these kernels are depicted in Figure~\ref{fig:Execution-latencies-of}.
It is clear that the execution latency does not have a linear relationship
with the $N_{MultipleHP-H-preld}$ and the execution latency might
increase as $N_{MultipleHP-H-preld}$ gets larger. Unfortunately,
even the largest $N_{MultipleHP-H-preld}$ ($5\times2^{15}$) used
in the experiments, and limited by the available FPGA resources, contains
only $4.7\%$ of the total number of all memory accesses. The best
performance achieved on $S5$ and $A10$ are both by executing kernel
\textsc{MultipleHP}-H-($5\times2^{13}$). Some improvements could
be made by overlapping the loading of the high touching frequency
points with the computing part, but not substantially. Overall, \textsc{MultipleHP}-H
is not gaining performance if the local memory size is not large enough
to hold most of the points with high touching frequency.

\paragraph{\textsc{MultipleHP-N}}

For kernel \textsc{MultipleHP}-N, the necessary data for each work
group are from nonconsecutive addresses and this affects the loading
section in achieving streaming mode, which is crucial  to fully use
the available theoretical bandwidth. Although executing more columns
per work group can reduce GMA, the value of $N_{MultipleHP-N-col}$
does not affect performance. The execution latency of \textsc{MultipleHP}-N
is affected by the kernel frequency, which is given in Table~\ref{tab:Resource-Multiple-n}.
We employ the kernel with the fastest execution latency to compare
with other methods, which is \textsc{MultipleHP}-N-$(1)$. 

\paragraph{\textsc{MultipleHP-R}}

The kernel execution latency and global memory occupancy during execution
on $A10$ are given in Table~\ref{tab:nc/wg=00003D16 64-1} as well.
When the value of $N_{lpoints/cc}$ is a power of 2, the execution
latency decreases as $N_{points/wi}$ increases. Although the occupancy
of loading operations drops, the values for the optimised kernels
decreases slower than that of the general kernels. The fastest variant
of \textsc{MultipleHP}-R in Table~\ref{tab:nc/wg=00003D16 64-1}
is \textsc{MultipleHP}-R-$(64,8)$. By adding the candidate detection,
the execution latency increases, however, faster than other \textsc{MultipleHP}
kernels. For kernel \textsc{MultipleHP}-R-$(16,4)$, the execution
latencies on a single $S5$ and $A10$ are $143ms$ and $120ms$,
respectively.

\paragraph{Overall Comparison}

\begin{figure}
\begin{centering}
\includegraphics[bb=0bp 15bp 576bp 280bp,clip,scale=0.45]{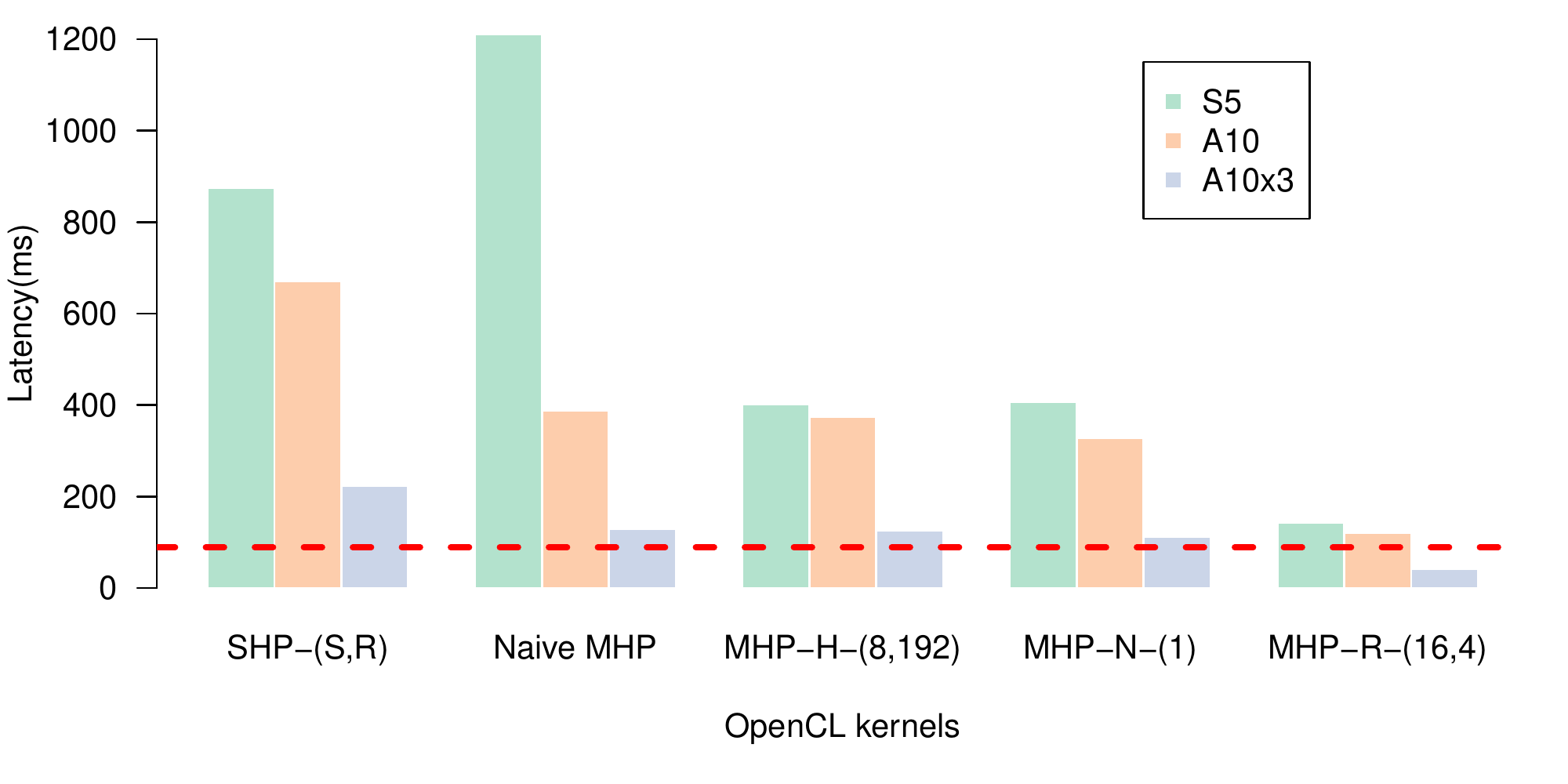}
\par\end{centering}
\caption{\label{fig:Execution-latency-of}Execution latency of proposed harmonic
summing methods with candidate detection on $A10$, where SHP represents
\textsc{SingleHP} and MHP represents \textsc{MultipleHP}}
\end{figure}

Based on the discussion above, the execution latency of each well-optimised
method with candidate detection is given in Figure~\ref{fig:Execution-latency-of},
and both types of FPGA devices are evaluated, where the red dashed
line is the current time limitation for the SKA harmonic summing.
We also evaluate a setting where three A10 FPGAs are used in parallel. 

Note that\textsc{, SingleHP-$(M,R)$} and $N_{paral}=16$ on $S5$
cost a large number of RAM blocks and cannot be compiled, hence, \textsc{SingleHP-$(S,R)$}
with $N_{paral}=8$ is used. The execution latency of \textsc{MultipleHP}-N-$(1)$
is faster than that of Na\text{\"i}ve \textsc{MultipleHP} and \textsc{MultipleHP}-H-$(8,192)$,
however, it is about 3x times slower than \textsc{MultipleHP}-R-$(16,4)$.
Except for Na\text{\"i}ve\textsc{ MultipleHP} on $S5$, all \textsc{MultipleHP}
kernels perform better than\textsc{ SingleHP}-$(S,R)$ with $N_{paral}=8$. 

Although the performance is improved by adopting \textsc{MultipleHP}
kernels, none of these kernels on a single $A10$ meets the requirement.
By installing three $A10$ FPGA cards, they can work in parallel by
processing three different half FOPs. The average execution latencies
of half FOP using three $A10$ cards are given in Figure~\ref{fig:Execution-latency-of}
as well. It can be seen that kernel \textsc{MultipleHP}-R on three
$A10$ cards is over 2x times faster than the required time limitation,
so three $A10$ cards can process the whole FOP while meeting the
requirements.

\subsubsection{\label{subsec:Comparison-with-CPU}Comparison with CPU and GPU}

We are now comparing the performance of the proposed kernels on GPU
(using adjusted OpenCL code) and CPU (using equivalent OpenMP implementations).
\textsc{SingleHP}-($M,\,$) is evaluated on $R7$ GPU, and the host
argument settings are the same as for the FPGA-based implementation.
The straightforward C code with OpenMP directives, using three levels
of \texttt{for} loops, which is the same as Algorithm~\ref{alg:General-Harmonic-summing-Algorit},
is evaluated on the $I7$ CPU using all four cores. The execution
latency of \textsc{SingleHP} using one core of $I7$ CPU ($I7-1C$)
is taken as the baseline and the speedups over it on other devices
are given in Table~\ref{tab:Parameters-of-testing-1}, where $I7-4C$
represents using four cores of the $I7$ CPU. It can be seen that
$R7$ performs best among these devices and it is about 3.6x times
faster than the $A10$ FPGA. The $R7$ has two major advantages over
$S5$ and $A10$: 1) operating frequency and 2) off-chip memory bandwidth.
Though the maximum frequency of $A10$ is higher than $R7$, the maximum
frequencies of the implemented kernels are less than $300MHz$ in
this work.

\begin{table}
\caption{\label{tab:Parameters-of-testing-1}Speedup of multi-core CPU, GPU,
and FPGA platforms over single core CPU in processing \textsc{SingleHP
}kernel including candidates detection }
\begin{centering}
{\footnotesize{}}%
\begin{tabular}{|c|c|c|}
\hline 
{\footnotesize{}Device} & {\footnotesize{}Execution latency($ms$)} & {\footnotesize{}Speedup over $I7-1C$}\tabularnewline
\hline 
\hline 
{\footnotesize{}$S5$} & {\footnotesize{}$875$} & {\footnotesize{}4.8}\tabularnewline
\hline 
{\footnotesize{}$A10$} & {\footnotesize{}$671$} & {\footnotesize{}6.2}\tabularnewline
\hline 
{\footnotesize{}$R7$} & {\footnotesize{}$119$} & \textbf{\footnotesize{}35.2}\tabularnewline
\hline 
{\footnotesize{}$I7-4C$} & {\footnotesize{}$1,100$} & {\footnotesize{}3.8}\tabularnewline
\hline 
{\footnotesize{}$I7-1C$} & {\footnotesize{}$4,174$} & {\footnotesize{}1}\tabularnewline
\hline 
\end{tabular}
\par\end{centering}{\footnotesize \par}
\centering{}{\small{}}{\small \par}
\end{table}

Regarding the \textsc{MultipleHP} kernels on GPU, a similar OpenCL
code as used for the FPGA kernels of Na\text{\"i}ve\textsc{ MultipleHP}
and \textsc{MultipleHP}-H are tested. The execution latencies of these
kernels are both over \textbf{$30$ }seconds, which are about a hundred
times slower than that of a single $A10$ FPGA. Because these two
variants are single work-item kernels, the GPU cannot parallelise
operations on multiple stream processors. For the fastest \textsc{MultipleHP}
kernel on $A10$, which is \textsc{MultipleHP}-R-$(64,8)$ (NDRange
kernel part), the execution latency of it (without candidates detection)
on $R7$ is $19.7ms$, and it is 3.7 times faster than achieved on
$A10$. After combining with the candidate detection, which is a single
work-item kernel, the performance drops as $N_{cand}$ increases.
When $N_{cand}=1$, the execution latency is $46.8ms$. However, when
$N_{cand}$ is increased to 200, the latency increases to $10$ seconds.
Since single work-item kernels on GPU cannot explore their performance
potential, we only compare the performance of NDRange kernels on FPGA
and GPU devices. 

Based on the above, an $R7$ is over 3.7 times faster than an $A10$
in executing the same NDRange kernels. Regarding the single work-item
kernels, GPU implementations cannot compete with FPGAs, being tens
to hundreds of times slower than FPGAs.

\subsection{Energy Dissipation and Power Consumption}

The execution latency is a significant performance criterion for the
harmonic-summing module. However, in the context of the pulsar search
engine in SKA1-MID, there will be over 2,000 beams that need to be
computed in parallel, which is constantly done for many years. As
a result, the power consumption is another essential criterion which
we investigate in this subsection.

To do so, we calculate the difference between the system power consumption
$P_{idle}$, including the acceleration device, in idle status and
the power consumption $P_{running}$ when the system is executing
the kernel. To make sure the value of $P_{running}$ is stable, each
kernel is launched hundreds of times using a loop, which takes several
minutes. 

The power consumption is measured using a plug-in power meter (Ego
smart socket ESS-AU). For the FPGA measurements, the calculated power
consumption is the value of using three $A10$ cards in one host.
The power consumption and energy dissipation of executing different
kernels are given in Table~\ref{tab:Energy-Dissipation-and}. The
energy cost is the dissipation of processing the input half FOP, and
the energy saving ratio is compared with the $I7-1C$. Since the execution
latencies of \textsc{MultipleHP} kernels with the single work-item
kernel (in Section~\ref{subsec:Comparison-with-CPU}) on GPU are
over ten times larger than those on FPGA, the \textsc{MultipleHP}
kernels with single work-item part are not compared with GPU. 

\begin{table}
\caption{\label{tab:Energy-Dissipation-and}Power consumption and energy dissipation
of FPGA, GPU, and CPU platforms (without candidate detection)}
\begin{centering}
{\footnotesize{}}
\par\end{centering}{\footnotesize \par}
\centering{}{\footnotesize{}}%
\begin{tabular}{|c|c|c|c|}
\hline 
\multirow{2}{*}{{\footnotesize{}Kernel-Setting (Device)}} & {\footnotesize{}Power} & {\footnotesize{}Energy} & \multirow{2}{*}{{\footnotesize{}Saving ratio}}\tabularnewline
 & {\footnotesize{}($watts$)} & {\footnotesize{}($Joules$)} & \tabularnewline
\hline 
\hline 
\textsc{\footnotesize{}SingleHP}{\footnotesize{}-($M,R$) ($A10\times3$)} & {\footnotesize{}23} & {\footnotesize{}3.36} & {\footnotesize{}19.9}\tabularnewline
\hline 
\textsc{\footnotesize{}SingleHP}{\footnotesize{}-($M,\,$) ($R7$)} & {\footnotesize{}65} & {\footnotesize{}8.9} & {\footnotesize{}7.5}\tabularnewline
\hline 
\textsc{\footnotesize{}SingleHP}{\footnotesize{}($I7-4C$)} & {\footnotesize{}43} & {\footnotesize{}47.85} & {\footnotesize{}1.4}\tabularnewline
\hline 
\textsc{\footnotesize{}SingleHP}{\footnotesize{}($I7-1C$)} & {\footnotesize{}16} & {\footnotesize{}66.8} & {\footnotesize{}1}\tabularnewline
\hline 
{\footnotesize{}Na\text{\"i}ve}\textsc{\footnotesize{} MultipleHP}{\footnotesize{}($A10\times3$)} & {\footnotesize{}7} & {\footnotesize{}0.91} & {\footnotesize{}73.4}\tabularnewline
\hline 
\textsc{\footnotesize{}MultipleHP}{\footnotesize{}-H ($A10\times3$)} & {\footnotesize{}10} & {\footnotesize{}1.75} & {\footnotesize{}38.2}\tabularnewline
\hline 
\textsc{\footnotesize{}MultipleHP}{\footnotesize{}-N ($A10\times3$)} & {\footnotesize{}14} & {\footnotesize{}1.11} & {\footnotesize{}60.0}\tabularnewline
\hline 
\textsc{\footnotesize{}MultipleHP}{\footnotesize{}-R ($R7$)} & {\footnotesize{}49} & {\footnotesize{}0.965} & {\footnotesize{}69.2}\tabularnewline
\hline 
\textsc{\footnotesize{}MultipleHP}{\footnotesize{}-R ($A10\times3$)} & {\footnotesize{}22} & {\footnotesize{}0.526} & \textbf{\footnotesize{}127.0}\tabularnewline
\hline 
\end{tabular}{\footnotesize \par}
\end{table}

Although the execution latency on $R7$ is faster than that of $A10$,
the energy dissipation of $R7$ is over 1.8 times higher than that
of three $A10$s. An interesting observation from Table~\ref{tab:Energy-Dissipation-and}
is that the power consumption of kernel \textsc{SingleHP}-($M,R$)
and \textsc{MultipleHP}-R on $A10$ are significantly higher than
other \textsc{MultipleHP} kernels on $A10$. The main reason is that
the used GMB of \textsc{SingleHP}-($M,R$) and \textsc{MultipleHP}-R
are optimised and much higher than other kernels. Streaming data between
off-chip memory and FPGA makes the power consumption of a kernel up
to 3 times higher than that of other \textsc{MultipleHP} kernels. 

In summary, it can be found that a single $R7$ needs over 2x times
more power than three $A10$ cards. Regarding the energy dissipation,
the cost of $R7$ is up to 2.6x times higher than three $A10$ cards
in executing the same kernels while providing similar performance.

\section{\label{sec:Conclusions}Conclusions}

In this paper, we investigated FPGA designs of one module of the SKA
pulsar search engine called harmonic-summing. OpenCL was chosen to
implement the proposed designs, and two types of FPGA cards (Intel
Stratix V and Arria 10 FPGAs) and a GPU card were employed for evaluation.
Two approaches of harmonic-summing were studied: 1) store intermediate
data in off-chip memory and 2) process the input signals directly
without storing intermediate data. For the second approach, since
a naive implementation does not provide good performance, two approaches
of preloading data were proposed and evaluated: 1) preloading points
that are touched most 2) preloading all necessary points that are
used to generate a chunk of output points. For the necessary points
approaches, the reorder of input signals is investigated as well. 

The extensive experimental evaluation demonstrated that kernels with
intermediate data storage perform worse than kernels without storing
intermediate data in both execution latency and power consumption.
A single FPGA can achieve 9.5x speedup over single-core CPU using
the general \textsc{SingleHP} method. By using three $A10$ FPGAs,
the NDRange \textsc{MultipleHP} kernels perform significantly better
than a single $R7$ GPU in power consumption, while only being slightly
slower regarding execution latency. To process the same amount of
data using the same OpenCL kernel, $R7$ GPU costs up to 2.6x times
more energy than three $A10$ FPGAs. This work shows that FPGA devices
can be a good solution for the SKA project for the processing parts
of the pulsar search pipeline.

\section*{Acknowledgment}

The authors acknowledge discussions with the TDT, a collaboration
between Manchester and Oxford Universities, and MPIfR Bonn and the
work benefitted from their collaboration. We would like to thank Petr
Dobias and Emmanuel Casseau from IRISA, University of Rennes 1.
We gratefully acknowledge that this research was financially supported by the 
SKA funding of the New Zealand government through the Ministry of Business, 
Innovation and Employment (MBIE).


\begin{thebibliography}{1}

\bibitem{canis2011legup}
Andrew Canis, Jongsok Choi, Mark Aldham, Victor Zhang, Ahmed
Kammoona, Jason H Anderson, Stephen Brown, and Tomasz Czajkowski.
Legup: high-level synthesis for fpga-based processor/
accelerator systems. In
\emph{Proceedings of the 19th ACM/SIGDA international symposium on Field 
programmable gate arrays}, pages 33-36. ACM, 2011.

\bibitem{carilli2004science}
Christopher Carilli  and Steve Rawlings,
Science with the Square Kilometer Array: motivation, key science projects, standards and assumptions,
\emph{arXiv preprint astro-ph/0409274}, 2004.

\bibitem{chen2012invited}
Doris Chen  and  Deshanand Singh,
Invited paper: Using OpenCL to evaluate the efficiency of CPUS, GPUS and FPGAS for information filtering,
\emph{In 22nd International Conference on Field Programmable Logic and Applications (FPL)},
5--12. IEEE, 2012.

\bibitem{czajkowski2012opencl}
Tomasz S Czajkowski,  Utku Aydonat, Dmitry Denisenko, John Freeman, Michael Kinsner, David Neto, Jason Wong, 
Peter Yiannacouras, and Deshanand P Singh.
From OpenCL to high-performance hardware on FPGAs.
\emph{In 22nd International Conference on Field Programmable Logic and Applications (FPL)},
531--534. IEEE, 2012.
  
\bibitem{de2007radio}
Ludovico De Souza, John D Bunton, Ducan Campbell-Wilson, Roger J
Cappallo, and Bart Kincaid.
A radio astronomy correlator optimized for the Xilinx Virtex-4 SX FPGA,
  author={,  and Bunton,  D and Campbell-Wilson, Ducan and Cappallo, Roger J and Kincaid, Bart},
\emph{In International Conference on Field Programmable Logic and Applications (FPL)},
pages 62--67, IEEE, 2007.

\bibitem{dewdney2009square}
Peter E Dewdney, Peter J Hall, Richard T Schilizzi, and T Joseph LW Lazio.
The square kilometre array.
\emph{Proceedings of the IEEE}, 97(8):1482--1496, 2009.

\bibitem{eguro2012fpgas}
Ken Eguro and Ramarathnam Venkatesan.
FPGAs for trusted cloud computing.
\emph{In 22nd International Conference on Field Programmable Logic and Applications (FPL)},
pages 63--70. IEEE, 2012.


\bibitem{fowers2014high}
Jeremy Fowers, Kalin Ovtcharov, Karin Strauss, Eric S Chung, and Greg Stitt.
A high memory bandwidth fpga accelerator for sparse matrix-vector multiplication.
\emph{In Field-Programmable Custom Computing
Machines (FCCM), 2014 IEEE 22nd Annual International Symposium on},
pages 36--43. IEEE, 2014.

\bibitem{khronosopencl}
Khronos OpenCLWorking Group et al. The opencl specification, version
1.0. 29, 8 december 2008.

\bibitem{halstead2011exploring}
Robert J Halstead, Jason Villarreal, and Walid Najjar.
Exploring irregular memory accesses on fpgas.
\emph{In Proceedings of the 1st Workshop
on Irregular Applications: Architectures and Algorithms},
pages 31--34. ACM, 2011.

\bibitem{hiba2012memory}
Antal Hiba, Zoltan Nagy, and Miklos Ruszinko.
Memory access optimization for computations on unstructured meshes,
\emph{In Cellular Nanoscale Networks and Their Applications (CNNA), 2012 13th International
Workshop on},
pages 1--5. IEEE, 2012.

\bibitem{huang2017hardware}
Sitao Huang, Gowthami Jayashri Manikandan, Anand Ramachandran,
Kyle Rupnow, W Hwu Wen-mei, and Deming Chen. Hardware
Acceleration of the Pair-HMM Algorithm for DNA Variant Calling.
\emph{In Proceedings of the 2017 ACM/SIGDA International Symposium on
Field-Programmable Gate Arrays},
pages 275--284. ACM, 2017.

\bibitem{altera2016openclpra}
Intel. Intel FPGA SDK OpenCL Best Pratices Guide, 2016.

\bibitem{jain2013linearizing}
Akanksha Jain and Calvin Lin. Linearizing irregular memory accesses
for improved correlated prefetching.
\emph{In Proceedings of the 46th Annual
IEEE/ACM International Symposium on Microarchitecture},
pages 247--259. ACM, 2013.

\bibitem{jang2011exploiting}
Byunghyun Jang, Dana Schaa, Perhaad Mistry, and David Kaeli. Exploiting
memory access patterns to improve memory performance in
data-parallel architectures.
\emph{IEEE Transactions on Parallel and Distributed Systems},
22(1):105--118, 2011.

\bibitem{leber2011high}
Christian Leber, Benjamin Geib, and Heiner Litz. High frequency
trading acceleration using fpgas.
\emph{In Field Programmable Logic and
Applications (FPL), 2011 International Conference on},
pages 317--322. IEEE, 2011.

\bibitem{mellor2001improving}
John Mellor-Crummey, David Whalley, and Ken Kennedy. Improving
memory hierarchy performance for irregular applications using data and
computation reorderings.
\emph{International Journal of Parallel Programming},
29(3):217--247, 2001.

\bibitem{parsons2009digital}
Aaron Parsons, Dan Werthimer, Donald Backer, Tim Bastian, Geoffrey
Bower, Walter Brisken, Henry Chen, Adam Deller, Terry Filiba, Dale
Gary, et al. Digital instrumentation for the radio astronomy community.
\emph{arXiv preprint arXiv:0904.1181}, 2009.

\bibitem{pavel2013algorithms}
Karas Pavel and Svoboda David. Algorithms for efficient computation of
convolution. In
\emph{Design and Architectures for Digital Signal Processing}
InTech, 2013.

\bibitem{putnam2014reconfigurable}
Andrew Putnam, Adrian M Caulfield, Eric S Chung, Derek Chiou,
Kypros Constantinides, John Demme, Hadi Esmaeilzadeh, Jeremy Fowers,
Gopi Prashanth Gopal, Jan Gray, et al. A reconfigurable fabric for
accelerating large-scale datacenter services. In
\emph{In Computer Architecture
(ISCA), 2014 ACM/IEEE 41st International Symposium on},
pages 13--24. IEEE, 2014.

\bibitem{ransom2002fourier}
Scott M Ransom, Stephen S Eikenberry, and John Middleditch. Fourier
techniques for very long astrophysical time-series analysis.
\emph{The Astronomical Journal}, 124(3):1788, 2002.

\bibitem{sanchez2005digital}
MA Sanchez, Mario Garrido, Marisa Lopez-Vallejo, Jesus Grajal, and
Carlos Lopez-Barrio. Digital channelised receivers on fpgas platforms.
In \emph{Radar Conference, 2005 IEEE International},
pages 816--821. IEEE, 2005.

\bibitem{sridharan2016accelerating}
Srikanth Sridharan, Paolo Durante, Christian Faerber, and Niko Neufeld.
Accelerating particle identification for high-speed data-filtering using
opencl on fpgas and other architectures. In
\emph{Field Programmable Logic
and Applications (FPL), 2016 26th International Conference on},
pages 1--7. IEEE, 2016.

\bibitem{tarafdar2017enabling}
Naif Tarafdar, Thomas Lin, Eric Fukuda, Hadi Bannazadeh, Alberto
Leon-Garcia, and Paul Chow. Enabling flexible network fpga clusters
in a heterogeneous cloud data center. In
\emph{Proceedings of the 2017
ACM/SIGDA International Symposium on Field-Programmable Gate
Arrays}, pages 237--246. ACM, 2017.

\bibitem{wang2016high}
Haomiao Wang, Ming Zhang, Prabu Thiagaraj, and Oliver Sinnen.
FPGA-based Acceleration of FDAS Module Using OpenCL. In
\emph{Field
Programmable Technology (FPT), 2016 International Conference on},
pages 53--60. IEEE, 2016.

\bibitem{wang2016fpga}
Haomiao Wang, Ming Zhang, Prabu Thiagaraj, and Oliver Sinnen.
FPGA-based Acceleration of FDAS Module Using OpenCL. In
\emph{Field
Programmable Technology (FPT), 2016 International Conference on},
pages 53--60. IEEE, 2016.

\bibitem{wang2015addressing}
Xu Wang, Linan Huang, Yongxin Zhu, Yipeng Zhou, Huwan Peng, and
Haifei Xiong. Addressing memory wall problem of graph computation
in reconfigurable system. In
\emph{High Performance Computing and
Communications (HPCC), 2015 IEEE 7th International Symposium on
Cyberspace Safety and Security (CSS), 2015 IEEE 12th International
Conferen on Embedded Software and Systems (ICESS), 2015 IEEE 17th
International Conference on},
pages 302--307. IEEE, 2015.

\bibitem{weinhardt1999memory}
Markus Weinhardt and Wayne Luk. Memory access optimization and
ram inference for pipeline vectorization. In
\emph{International Conference on Field Programmable Logic and Applications (FPL)},
pages 61--70. Springer, 1999.

\bibitem{weinhardt2001memory}
Markus Weinhardt and Wayne Luk. Memory access optimisation
for reconfigurable systems.
\emph{IEEE Proceedings-Computers and Digital
Techniques}, 148(3):105--112, 2001.

\bibitem{yang2013optimizing}
Hsin-Jung Yang, Kermin Fleming, Michael Adler, and Joel Emer.
Optimizing under abstraction: Using prefetching to improve fpga performance.
In \emph{Field Programmable Logic and Applications (FPL), 2013
23rd International Conference on}, pages 1--8. IEEE, 2013.
\end{thebibliography}

\end{document}